\newif\ifsubmission
\submissionfalse

\ifsubmission
\documentclass[conference,compsoc]{IEEEtran}
\else
\documentclass[letterpaper,11pt]{article}
\usepackage[margin=1in]{geometry}
\usepackage[colorlinks,citecolor=blue]{hyperref}
\fi
\usepackage{amsmath,amscd}
\usepackage{amsfonts}
\usepackage{amsthm}

\usepackage{cleveref}
\usepackage{bussproofs}
\usepackage{nicefrac}
\usepackage{siunitx}
\usepackage{array,framed}
\usepackage{collectbox}
\usepackage[markup=underlined]{changes}
\usepackage{hyperref}
\usepackage{booktabs}
\usepackage{
  color,
  float,
  epsfig,
  wrapfig,
  graphics,
  graphicx,
  subcaption
}
\usepackage[font=footnotesize]{caption}

\usepackage[switch]{lineno}
\usepackage{lipsum}%

\usepackage{textcomp} %
\usepackage{latexsym,fancyhdr,url}
\usepackage{enumerate}
\usepackage{enumitem}
\usepackage{soul}

\usepackage{graphics}
\usepackage{xparse} %
\usepackage{xspace}
\usepackage{multirow}
\usepackage{csvsimple}
\usepackage{balance}
\usepackage[symbol]{footmisc}
\usepackage{
  tikz,
  pgfplots,
  pgfplotstable
}
\usetikzlibrary{
  shapes.geometric,
  arrows,
  external,
  pgfplots.groupplots,
  matrix
}
\usepackage{tikz}
\usetikzlibrary{arrows.meta, shapes, positioning}
\usepackage{etoolbox,xspace}

\pgfplotsset{compat=1.9}

\usepackage{mathtools,}

\DeclareMathAlphabet{\mathcal}{OMS}{cmsy}{m}{n}

\DeclareGraphicsExtensions{%
    .png,.PNG,%
    .pdf,.PDF,%
    .jpg,.mps,.jpeg,.jbig2,.jb2,.JPG,.JPEG,.JBIG2,.JB2}

\usepackage{graphicx,subcaption,booktabs,multirow,wrapfig,svg}
\pgfplotsset{compat=1.9}
\usetikzlibrary{arrows.meta,shapes,positioning,calc,matrix,external,pgfplots.groupplots}

\usepackage[noend]{algpseudocode}

\usepackage{enumitem}

\usepackage{xspace,nicefrac,soul,balance,collectbox,color,textcomp}

\usepackage[nocompress]{cite}
\usepackage{float}

\usepackage{url}
\newtheorem{theorem}{Theorem}
\usepackage{tabularx}
\usepackage{algorithm}
\usepackage{algpseudocode}
\usepackage{titlesec}
\setlist{leftmargin=3.5mm}
\newtheorem{formula}{Formula} %
\crefname{formula}{Formula}{Formula}
     \titlespacing*{\subsection}{0pt}{1ex}{.5ex}
     \titlespacing*{\section}{0pt}{1ex}{.5ex}
     \titlespacing*{\subsubsection}{0pt}{1ex}{.5ex}

\usepackage{xparse}
\newcommand{\bnm}{\begin{newmath}}
\newcommand{\enm}{\end{newmath}}

\newcommand{\bea}{\begin{eqnarray*}}%
\newcommand{\eea}{\end{eqnarray*}}%

\newcommand{\bne}{\begin{newequation}}
\newcommand{\ene}{\end{newequation}}

\newcommand{\bal}{\begin{newalign}}
\newcommand{\eal}{\end{newalign}}

\newenvironment{newalign}{\begin{align}%
\setlength{\abovedisplayskip}{4pt}%
\setlength{\belowdisplayskip}{4pt}%
\setlength{\abovedisplayshortskip}{6pt}%
\setlength{\belowdisplayshortskip}{6pt} }{\end{align}}

\newenvironment{newmath}{\begin{displaymath}%
\setlength{\abovedisplayskip}{4pt}%
\setlength{\belowdisplayskip}{4pt}%
\setlength{\abovedisplayshortskip}{6pt}%
\setlength{\belowdisplayshortskip}{6pt} }{\end{displaymath}}

\newenvironment{newequation}{\begin{equation}%
\setlength{\abovedisplayskip}{4pt}%
\setlength{\belowdisplayskip}{4pt}%
\setlength{\abovedisplayshortskip}{6pt}%
\setlength{\belowdisplayshortskip}{6pt} }{\end{equation}}

\newcounter{mytable}
\def\mytable{\begin{minipage}{\textwidth}\centering\refstepcounter{mytable}}
\def\endmytable{\end{minipage}}

\newcounter{myfig}
\def\myfig{\begin{centering}\refstepcounter{myfig}}
\def\endmyfig{\end{centering}}

\newlength{\saveparindent}
\newlength{\saveparskip}
\newcommand{\E}{{\rm I\kern-.3em E}}

\renewcommand{\eqref}[1]{\mbox{Equation~(\ref{#1})}}

\newcommand{\N}{\ensuremath{\mathbb{N}}}
\newcommand{\NN}{\ensuremath{\mathbb{N}}}

\newcommand{\FF}{\ensuremath{\mathbb{F}}}

\def \part {part}

\renewcommand{\paragraph}[1]{\vspace*{6pt}\noindent\textbf{#1}\;}

\def \blackslug{\hbox{\hskip 1pt \vrule width 4pt height 8pt
    depth 1.5pt \hskip 1pt}}
\def \qed{\quad\blackslug\lower 8.5pt\null\par}

\newcounter{mynote}[section]

\newcommand\ignore[1]{}

\newcounter{rcnote}[section]

\newcounter{mrnote}[section]

\newcounter{fknote}[section]

\newcounter{anote}[section]

\DeclareMathSymbol{\mlq}{\mathord}{operators}{``}
\DeclareMathSymbol{\mrq}{\mathord}{operators}{`'}

\newcommand{\rhf}[2]{R_{f, \gamma}}

\DeclareDocumentCommand{\edist}{o o}{
  \ensuremath{
    \IfNoValueTF{#1}{{d}}{{\sf d}(#1,#2)}
  }
}

\newcommand{\olrk}[1]{\ifx\nursymbol#1\else\!\!\mskip4.5mu plus 0.5mu\left(\mskip0.5mu plus0.5mu #1\mskip1.5mu plus0.5mu \right)\fi}

\NewDocumentCommand{\indseq}{ O{1} O{r} }{{#1}\ldots {#2}}

\newcommand{\forallrmv}{\xspace$\forall-${\sc Red}\xspace}
\newcommand{\shortsection}[1]{\smallskip\noindent{\bf #1}}
\newcommand{\xres}{{$\exists$-\textsc{Res}\xspace}}
\newcommand{\ured}{{$\forall$-\textsc{Red}\xspace}}
\newcommand{\qres}{{\textsc{Q-Res}}\xspace}

\newcommand{\Prv}{{\ensuremath{\mathcal{P}}}\xspace}

\newcommand{\zkqres}{\textsc{ZKQRES}\xspace}
\newcommand{\zkws}{\textsc{ZKWS}\xspace}

\newcommand{\Ver}{{\ensuremath{\mathcal{V}}}\xspace}

\newcommand{\Dep}{\lambda}
\newcommand{\lits}{\mathcal{L}}
\newcommand{\var}{\mathcal{X}}
\newcommand{\lexists}{\mathcal{L}_{\exists}}
\newcommand{\lforall}{\mathcal{L}_{\forall}}
 \newcommand{\varexists}{\mathcal{X}_{\exists}}
\newcommand{\varforall}{\mathcal{X}_{\forall}}
\newcommand{\varaux}{\mathcal{X}_{\sf aux}}
\newcommand{\laux}{\mathcal{L}_{\sf aux}}
\newcommand{\polynomial}{\mathcal{P}}
\newcommand{\ordlength}{k}
\newcommand{\flength}{{L}}
\newcommand{\xzero}{{x_0}}
\newcommand{\xone}{{x_1}}
\newcommand{\QBFEVAL}{QBFEVAL}
\newcommand{\getorder}{\mathsf{\xi}}

\newcommand{\order}{{\sf order}\xspace}
\newcommand{\dep}{{\Dep}}

\newcommand{\xdownarrow}[1]{%
  {\left\downarrow\vbox to #1{}\right.\kern-\nulldelimiterspace}
}

\def\Tau{{\rm T}}

\newenvironment{nffunc}[1]{
 \small
  \begin{framed}
    \begin{minipage}{\linewidth}
\vspace{-3pt}
       {\begin{center}\underline{\textbf{Functionality} #1}\end{center}}
}{
\vspace{-3pt}
    \end{minipage}
  \end{framed}
}

\newenvironment{nfprot}[1]{
  \small
  \begin{framed}
    \begin{minipage}{\linewidth}
\vspace{-3pt}
      {\begin{center}\underline{\textbf{Protocol} #1}\end{center}}
}{
\vspace{-3pt}
    \end{minipage}
  \end{framed}
}

\newcommand{\Func}[1][\relax]{\ensuremath{\mathcal{F}_\mathsf{#1}}\xspace}
\newcommand{\Prot}[1][\relax]{\ensuremath{\Pi_\mathsf{#1}}\xspace}

\title{Towards Practical Zero‑Knowledge Proof for PSPACE}

\begin{document}

\ifsubmission
\else
\author{
Ashwin Karthikeyan \\
University of Toronto\\
\texttt{\href{mailto:ashwin@cs.toronto.edu}{ashwin@cs.toronto.edu}}
\and
Hengyu Liu \\
University of Illinois Urbana–Champaign\\
\texttt{\href{mailto:hengyu2@illinois.edu}{hengyu2@illinois.edu}}
\and
Kuldeep S.\ Meel\\
University of Toronto, Georgia Institute of Technology\\
\texttt{\href{mailto:meel@cs.toronto.edu}{meel@cs.toronto.edu}}
\and
Ning Luo\footnote{Ning Luo is the corresponding author -
  \href{mailto:nl27@illinois.edu}{nl27@illinois.edu}}\\
University of Illinois Urbana–Champaign\\
\texttt{\href{mailto:nl27@illinois.edu}{nl27@illinois.edu}}
}
\fi
\date{}
\maketitle

\vspace{-2em}

\begin{abstract}
Efficient zero-knowledge proofs (ZKPs) have been restricted to NP statements so far, whereas they exist for all statements in PSPACE.
This work presents the first practical zero-knowledge (ZK) protocols for PSPACE-complete statements by enabling ZK proofs of QBF (Quantified Boolean Formula) evaluation. The core idea is to validate quantified resolution proofs (\qres) in ZK. We develop an efficient polynomial encoding of \qres proofs, enabling proof validation through low-overhead arithmetic checks. We also design a ZK protocol to prove knowledge of a winning strategy related to the QBF, which is often equally important in practice.
We implement our protocols and evaluate them on QBFEVAL.
The results show that our protocols can verify 72\% of QBF evaluations via \qres proof and 82\% of instances' winning strategies within 100 seconds, for instances where such proofs or strategies can be obtained.

\end{abstract}

\section{Introduction}
\label{sec:intro}
Zero-knowledge proofs (ZKPs) enable one party (the prover) to prove to another party (the verifier) the validity of a statement without revealing any information beyond the statement\cite{FOCS:GolMicWig86}. Such power makes ZKP a fundamental tool for achieving privacy and verifiability in many computational settings, particularly since recent advances in efficient ZKP constructions have significantly improved their practicality. We have witnessed ZKP's deployment in many real-world applications, including authentication, scalable blockchain protocols, and privacy-preserving machine learning\cite{EC:Groth16,EC:GGPR13,SP:PHGR13,C:BCGTV13,USENIX:BCTV14,NDSS:WSRBW15,C:HuMohRos15,EC:MohRosSca17,CCS:AHIV17,AC:BCGJM18,TCC:BHRRS20,CCS:HeaKol20,C:BHRRS21,heath2021zero,CCS:FKLOWW21}. 
Most efforts to make ZKPs efficient have focused on NP statements, leaving statements of practical interest in PSPACE under-investigated.

We consider a simple two-player game between a prover and a verifier on a $2 \times 2$ grid where the coordinate of the upper left cell is $(0,0)$, the upper right cell is $(1,0)$, and the bottom right cell is $(1,1)$. The prover starts at position $(0,0)$ and aims to reach the goal at $(1,1)$ in exactly two steps. The verifier may block either cell $(0,1)$ or $(1,0)$, but only after observing the prover’s first move. The prover moves first, choosing either right or down, and then, after the verifier selects a cell to block, makes a second move. The prover wins if she reaches $(1,1)$ without entering the blocked cell after the verifier blocked it. The goal of the prover is to convince the verifier that she knows a strategy to reach $(1,1)$ {\emph without revealing the strategy} itself. 

The interaction of the game can be encoded as a quantified Boolean formula (QBF):
\[
\exists \xzero \; \forall y \; \exists \xone. \; \texttt{Reachable}(\xone, y, \xzero)
\]
Here, $y \in \{0,1\}$ denotes the verifier's blocking choice: $y = 1$ corresponds to blocking cell $(0,1)$, and $y = 0$ corresponds to blocking $(1,0)$. The variables $\xzero$ and $\xone$  represent the prover’s first and second moves, respectively, with $0$ indicating a horizontal move and $1$ indicating a vertical move. The predicate $\texttt{Reachable}(\xone, y, \xzero)$ evaluates to true if the prover successfully reaches $(1,1)$ without entering the blocked cell.
For example, $\texttt{Reachable}(\xone = 1, y = 0, \xzero = 0) = 1$, since the prover first moves right to $(1,0)$, then the verifier blocks $(1,0)$, and then the prover moves down to reach $(1,1)$. The goal is reached successfully as the prover reached $(1,1)$ and did not enter the blocked cell $(1,0)$ after the verifier blocked it.

The QBF formula evaluates to true, as the prover has a strategy ($\xzero = 0, \xone=1$) that guarantees reaching the goal regardless of the verifier's blocking choice. This can be easily verified since the verifier’s blocking action occurs only after the prover moves to $(1,0)$, and therefore, the blocked cell has no effect on the chosen path. However, in more complex winning configurations, determining whether the prover has a winning strategy is nontrivial. On the other hand, most such position-based games can be encoded as QBF instances, where the existence of a winning strategy corresponds to the truth of the QBF.

QBF evaluation is PSPACE-complete and extends propositional logic by allowing both existential ($\exists$) and universal ($\forall$) quantification over Boolean variables. The inclusion of quantified variables enables the modeling of uncertainty, such as adversarial or environmental actions. This level of expressiveness exceeds that of NP and naturally arises in many practical domains, including software verification, circuit synthesis, and AI robustness.

Establishing ZKPs for QBFs has both theoretical value and practical significance. Due to the expressiveness of QBFs, ZKPs of QBFs can enable privacy-preserving proofs for a broad class of applications currently beyond the scope of existing ZKP frameworks. Below, we highlight the following use cases that are representative of this gap, and we will also illustrate this in our evaluation (see~\Cref{sec:eval}):

\noindent
 \underline{Partial Equivalence Checking (PEC)}~\cite{gitina2013equivalence}: A hardware designer wants to convince the customer that their opaque and partially specified implementation of a combinational circuit can still be completed into a full design that is functionally equivalent to the customer’s specified target for future integration.

\noindent
\underline{Conformant Planning (C-PLAN)}~\cite{egly2017conformant, rintanen2007asymptotically}: An AI service provider convinces its client to purchase a plan comprising a sequence of actions that transitions from initial states to a goal-satisfying state, regardless of the initial state or non-deterministic behavior of the environment. 

\noindent
\underline{Black Box Checking (BBC)}~\cite{Peled1999}: A white-hat hacker proves the existence of software bugs that are independent of the behaviors of unknown structures and modules.

These applications ask the verifier to trust not only a negative certificate of infeasibility but also a constructive guarantee of capability. For a winning strategy, for example, an executable conformant plan for C-PLAN or a bug-exhibiting trace for BBC, will turn the satisfiability of a QBF into an operational certificate, whereas UNSAT proofs cannot certify how to act or what is the strategy to win. This work makes such capability attestable in zero knowledge, we prove the existence of strategies without disclosing them.

In this work, we propose efficient ZKPs for QBF evaluation, which make ZKPs for other PSPACE statements tractable and thereby address the aforementioned challenges. In~\Cref{sec:eval}, we further elaborate our protocols' capabilities on {PEC}, {C-PLAN}, and {BBC} through evaluations on real-world benchmarks. 

\paragraph{Key challenges.}
The theoretical feasibility of constructing ZKPs for PSPACE languages follows from the classical result that $\mathsf{IP} = \mathsf{PSPACE}$~\cite{ipequalpspace, impagliazzo1987direct}, where the evaluation of a QBF is encoded as an interactive proof using the sumcheck protocol. A generic transformation then converts this interactive proof into a zero-knowledge protocol~\cite{ben1990everything}. However, this construction requires evaluating high-degree multivariate polynomials over large fields, with degrees scaling linearly with the number of variables. As a result, the approach incurs substantial computational overhead and is not suitable for real-world QBF instances. Designing efficient and scalable ZKPs for PSPACE-complete problems remains an open and largely unexplored direction.

Not only the truth value of a QBF but also the \emph{winning strategy} is of interest in practice. A winning strategy is a concrete example that describes how to assign existential (universal) quantified variables in response to universal (existential) quantified ones. In the grid game example, a winning strategy corresponds to a sequence of moves that ensures success regardless of the verifier’s responses. More generally, the winning strategy of a QBF encodes a functional dependency of existential (universal) quantified variables on the universal (existential) quantified variables, which can be represented via Skolem functions (Skolemization)~\cite{benedetti2005evaluating}. Existing ZKP protocols that only prove knowledge of a QBF’s evaluation do not take into consideration the knowledge of such a concrete winning strategy. On the other hand,  the knowledge of such a winning strategy is not only the object of verification in real-world applications, but also can improve the performance of the ZKP for QBF evaluation as the extended witness.

\paragraph{This work.} We design and implement a novel, efficient ZKP for the evaluation of public QBFs. Our protocol can be used directly to efficiently prove knowledge of any statements in PSPACE once the statement has been reduced to prove the evaluation of QBF. We also introduce a ZKP protocol for proving knowledge of a winning strategy for a given QBF.  In fact, we have demonstrated that with the aid of winning strategies, the resulting ZKP can be highly efficient for some instances, allowing us to reduce the proving time by 200X\footnote{It is not necessary for the winning strategy to improve the proving time as the size of the winning strategy can be significantly larger than the size of \qres proof.}.

Our first protocol leverages quantified resolution proofs (\qres) as an additional input from the prover, making ZKP for QBF evaluation and, therefore, PSPACE practical. 
\qres of a QBF consists of deriving clauses according to two rules: $\exists$-resolution ($\exists$-RES) and $\forall$-reduction ($\forall$-RED). The proof ends up with a derived empty clause, demonstrating that the formula is false. Although a \qres of a QBF can theoretically be exponentially large, for many QBFs encoding real-world problems, the resulting \qres proofs are of reasonable size. Modern QBF solvers can generate such proofs. 

To further make ZK for PSPACE practical, we propose a ZKP that enables the prover to prove her knowledge of a winning strategy rather than merely the evaluation of the QBF. 
If the prover is willing to disclose the size of this winning strategy, we find that the ZKP for PSPACE can be efficiently reduced to ZKP for validity (when the original QBF is true) or unsatisfiability (when the original QBF is false), with only a small overhead in verifying the correct dependency in the winning strategies in ZKP. Furthermore, ZKP for unsatisfiability and validity has been studied, with highly optimized protocols readily available. By leveraging these well-established techniques, our approach strikes a balance between efficiency and practicality, making it suitable for large-scale applications.

We improve the implementation of our protocols by analyzing the \qres proof structure in the context of ZKP. 
We propose a hierarchical encoding scheme that groups clauses by their size, improving runtime performance by approximately half. 
\ifsubmission
\else
\begin{figure}
    \centering
    \includegraphics[width=0.8\linewidth]{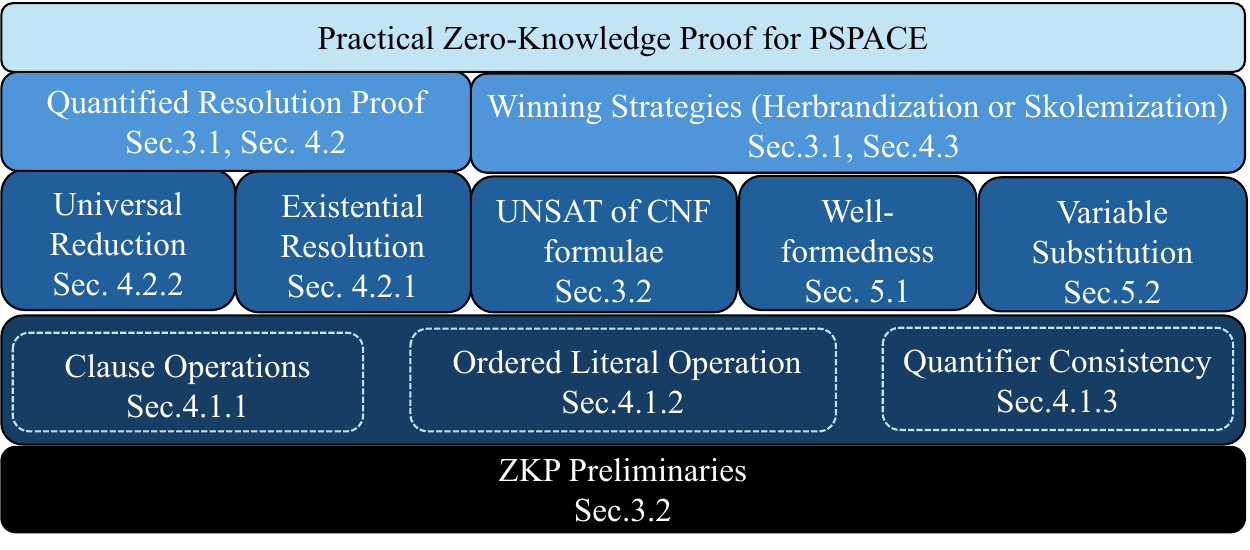}
    \caption{{\bf Paper roadmap.} We propose two practical approaches for proving QBF evaluation in ZK: one based on \qres proofs and the other based on winning strategies. We implement and evaluate both approaches, with results presented in~\Cref{sec:eval}.
}\vspace{-3mm}
    \label{fig:enter-label}
\end{figure}
\fi

\paragraph{Our Contribution.}
We present the first practical ZKP for the evaluation of a QBF. Our work expands ZKPs for PSPACE-complete problems and their applicability to real-world scenarios. \\
Our paper makes the following contribution:
\begin{itemize}
    \item We introduce a quantifier-encoding scheme that enables the verification of both Q-Resolution and Q-Cube-Resolution proofs, and we design a protocol to check that a private CNF (sub)formula encodes a private winning strategy by verifying its derivation from a private And-Inverter Graph (AIG).
    \item We develop a novel and efficient ZKP protocol for proving the evaluation of QBF by synergizing the advances in QBF reasoning and ZKP. Our approach provides an efficient way to encode the QBF and \qres using polynomials and perform validity checking of \qres proofs. 
    \item We enable the prover in ZKP to prove not only the evaluation of the QBF, but also the knowledge of the winning strategies. With the winning strategy, ZKP for QBF evaluation can be reduced to ZKP for UNSAT by revealing the size of the winning strategy. 
    \item We implement our protocols and evaluate them on QBFEVAL, a well-established benchmark suite for QBF solvers, and with instances derived from real-world problems. For QBFEVAL'07,  Out of 392 false QBFs for which we obtain \qres proofs or winning strategies, our protocols can verify QBF evaluations for about 82\% of instances in 100 seconds (see~\Cref{fig:verified_dist}). Instances from \textsc{PEC}, \textsc{C-Plan}, and \textsc{BBC} are verified within 300, 1,200, and 200 seconds, respectively.

    \item We present a highly optimized implementation of our protocols. To enhance efficiency, we also introduce a batching scheme that groups clauses into buckets of similar width, thereby minimizing the padding overhead. This optimization reduces the runtime by approximately 50\%.
\end{itemize}

Our implementation can be found at \href{https://github.com/PP-FM/zkqbf-suite}{https://github.com/PP-FM/zkqbf-suite}.

\ifsubmission
\else
\begin{figure}[h]
    \centering
    \includegraphics[width=0.8\linewidth]{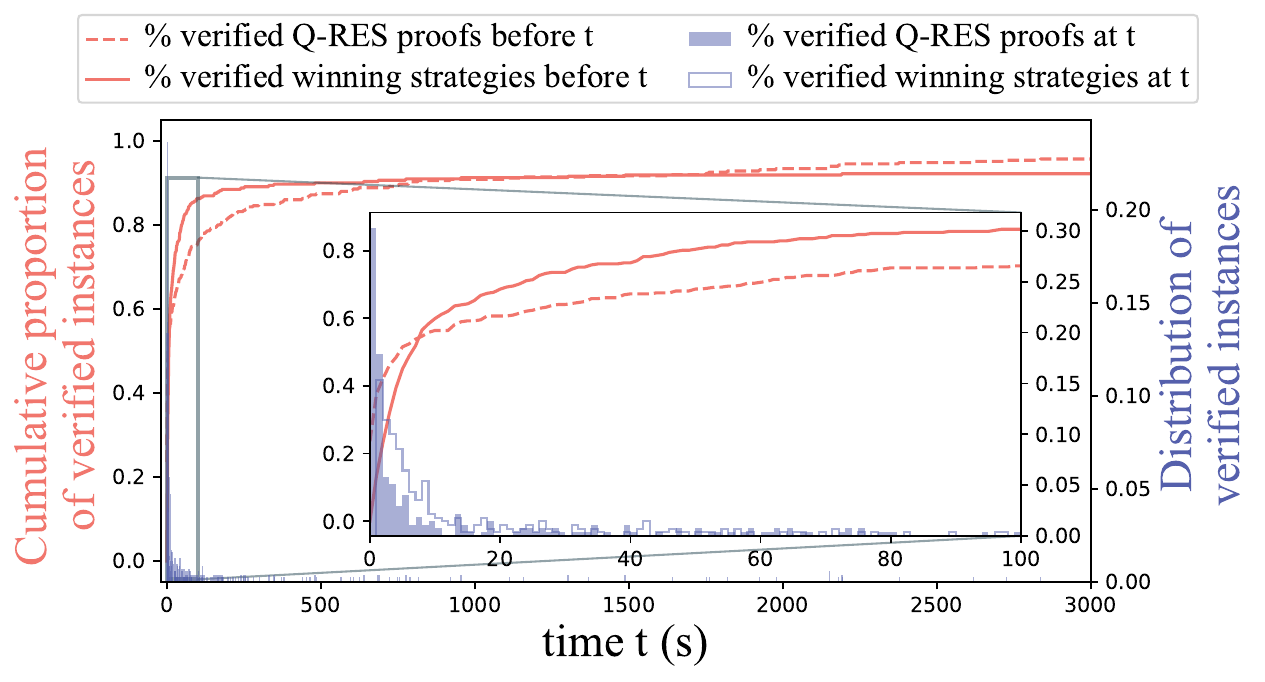}
    \caption{\footnotesize {\bf Our protocols' evaluation against QBFEVAL.} We present the cumulative fraction of QBF instances successfully verified by our protocols via Q-RES proofs and winning strategies within a given time threshold (left $Y$-axis), as well as the fraction verified around each time point (right $Y$-axis). Our protocols can verify 72\% of QBFs' evaluations via \qres proof and 82\% of instances' winning strategies within 100 seconds, for instances where such proofs or strategies can be obtained. See~\Cref{sec:eval} for details.}
    \label{fig:verified_dist}
\end{figure}
\fi

\subsection{Related Work}
\noindent
{\bf Proof systems for QBF.}
Proof systems for complexity classes beyond NP remain an active area of research, addressing both foundational questions in proof complexity and the development of practical solvers. Early work by Jussila, Sinz, and Biere~\cite{jussila2006extended, sinz2006extended} showed that extended resolution proofs can be systematically derived from Binary Decision Diagram (BDD) operations. This insight enabled proof-producing SAT and QBF solvers based on symbolic representations.
The primary proof systems for quantified Boolean formulas (QBFs) include Q- and QU-resolution, long-distance Q-resolution, $\forall$Exp+Res, IR-calc (instantiation and resolution calculus), IRM-calc (IR with merging), and merge resolution~\cite{Beyersdorff14, Beyersdorff19, Beyersdorff20, Beyersdorff2015}. Among these, IR-calc and IRM-calc provide a unified framework that captures both CDCL-style and expansion-based reasoning, and they support efficient extraction of universal strategies~\cite{Beyersdorff14}. The complexity of strategy extraction and its connection to proof size has also been studied extensively~\cite{BeyersdorffHardness18, Beyersdorff2015}.
Frege systems and their circuit-augmented variants~\cite{fregesystems} are known to simulate nearly all existing clausal and expansion-based QBF proof systems, offering a standard format for expressing complex proofs.
Beyond classical logics, interactive proofs provide an additional dimension of expressiveness and efficiency. The IP = PSPACE theorem enables the verification of PSPACE-complete languages using interactive protocols with polynomial-time verifiers~\cite{Das2017, CzernerEK24}.

We also acknowledge concurrent work by Kolesar et al. \cite{crepe} that targets ZK Protocols for PSPACE-Complete problems through proving the equivalence of two regular expressions.

\noindent
{\bf ZKPs beyond NP.} 
A classical result established in the 1980s shows that any language in the class IP admits a zero-knowledge proof, via a generic transformation from interactive proofs to ZKP~\cite{ben1990everything}. This foundational result ensures that all PSPACE-complete problems, including QBF, are in ZK. However, the transformation is not optimized for practical use and does not yield efficient or scalable ZKP protocols.
Recent work has advanced practical ZKPs for specific logical properties such as unsatisfiability and validity~\cite{zkunsat, luick2024zksmt, laufer2024zkpi}. ZKUNSAT~\cite{zkunsat} and ZKSMT~\cite{luick2024zksmt} implement zero-knowledge protocols for certifying unsatisfiability of propositional and SMT formulas, respectively. However, both frameworks lack support for quantifiers and therefore cannot address problems in PSPACE. A different direction is taken by zkPi~\cite{laufer2024zkpi}, which encodes formal proofs from interactive theorem provers, primarily Lean. zkPi accommodates a broad range of logical constructs, including algebraic data types, lambda calculus, and inductive reasoning. In contrast, our work targets scalable and efficient zero-knowledge protocols tailored explicitly for large QBF instances, and does not rely on proof traces from external theorem provers.
Recent developments in ZK extend to settings beyond NP using multi-prover interactive proofs (MIP$^*$) and protocols for NEXP~\cite{STOC2024Mastel, STOC2018Chiesa, iacr2016Sasson}. These systems rely on multiple provers and are thus orthogonal to the single-prover setting relevant to PSPACE. Moreover, the complexity-theoretic result QIP = PSPACE~\cite{QIP} establishes the equivalence between quantum interactive proofs and PSPACE, highlighting that PSPACE languages admit interactive proofs with quantum verifiers.

\section{A motivating example.}
Continuing with the grid game example, the prover can establish the truth of the original QBF by demonstrating that its negation is false. For clarity, we illustrate this using a simplified predicate, denoted as $\neg \texttt{Reachable}$. The goal of the prover is to falsify the following QBF:

\vspace{-1mm}
\begin{formula}
\begin{align}
\forall \xzero \; \exists y \; \forall \xone. \quad
&(\xzero \lor y \lor \xone) 
\land (\xzero \lor \neg y \lor \neg \xone)  \nonumber\\
&\underbrace{\land (\neg \xzero \lor y \lor \neg \xone) 
\land (\neg \xzero \lor \neg y \lor \xone)}_{\neg \texttt{Reachable}}
\end{align} 
\label{eq:qbf}
\end{formula}
\vspace{-4mm}
\noindent
{\bf $\forall$-Reduction (\ured):} 
Consider the clause $(\xzero \lor y \lor  \xone)$ under the quantifier prefix $\forall \xzero \; \exists y \; \forall \xone$ in the example. Since the existential variable $y$ appears before $\xone$ in the quantifier prefix, it cannot depend on $\xone$. Therefore, we can safely remove $\xone$ from the first clause, yielding $C_0 = (\xzero \lor y)$. This rule of deduction is called $\forall$-reduction. Using the same rule, we can also have $C_1 = (\neg \xzero \lor  y)$ from the third clause, and $C_2 = (\neg \xzero \lor \neg y)$ from the fourth clause. 

\noindent
{\bf $\exists$-Resolution (\xres):}  
Given the clauses $C_1$ and $C_2$, we can resolve on the existential variable $y$ to obtain $C_3 = (\neg \xzero)$. The derived clause $C_3$ is valid under the same quantifier prefix if the original QBF is valid and $C_3$ does not contain both a literal and its negation ($C_3$ is non-tautological). This rule of deduction is called the $\exists$-resolution.

We can apply \forallrmv again to $(\lnot \xzero)$ in the end, as $\xzero$ is the only universal variable in the clause. We obtain the empty clause $\bot$, which falsifies the original QBF. Such a process leading to the empty clauses by applying \ured{} and \xres{} is called a Q-resolution (\qres) proof. 
\begin{figure}[h]
\centering

\begin{prooftree}
\AxiomC{$(\neg \xzero \lor y \lor \neg \xone)$}
\RightLabel{\scriptsize\bf  $\forall$-Red on $\xone$}
\UnaryInfC{$(\neg \xzero \lor y)$}

\AxiomC{$(\neg \xzero \lor \neg y \lor \xone)$}
\RightLabel{\scriptsize\bf  $\forall$-Red on $\xone$}
\UnaryInfC{$(\neg \xzero \lor \neg y)$}
\RightLabel{\scriptsize\bf  Res on $y$}
\BinaryInfC{$(\neg \xzero)$}

\RightLabel{\scriptsize \bf $\forall$-Red on $\xzero$}
\UnaryInfC{$\bot$}
\end{prooftree}
\caption{{\bf The \qres proof for falsifying Formula~\ref{eq:qbf}.} Clauses are deducted via $\forall$-Red and \xres.  $\forall$-reduction removes universal literals from a clause when permitted by the quantifier prefix. \xres{} combines two clauses by eliminating the complementary literals. The deducted empty clause indicates that the evaluation of Formula~\ref{eq:qbf} is false.} \vspace{-2.5mm}
\label{fig:q-res}
\end{figure}

\noindent
To enable practical ZKP for QBF evaluation via \qres, the following functionalities should be efficiently supported to arithmetize the verification of the \qres proof.
\begin{itemize}
    \item[-] Verification of the quantifier type (universal or existential quantified) of each variable, and consistency of quantifiers of variables across proof steps;
    \item[-] Validation of the quantifier-induced order of literals to ensure the soundness of universal reduction;
    \item[-] Efficient checking of both existential resolution (\xres{}) and universal reduction (\ured{}) steps.
\end{itemize}

To achieve this, we first maintain two sets $\lforall$ and $\lexists$ for existential and universal quantified literals, respectively. 
Each literal is represented by a binary string where the first bit encodes the sign (0 for negation) and the remaining bits denote the variable order in binary. 
For example, $\neg \xone$ is encoded as \texttt{010} and $\xone$ as \texttt{110} given the quantifier prefix $\forall \xzero \exists y \forall \xone$. These binary strings admit dual interpretations: 1) as field elements in $\FF_{2^K}$ (for $K$-bit encodings, and 2) integer values through binary-to-decimal conversion.

Clauses are encoded via polynomials in $\FF_{2^k}[X]$ whose roots correspond to literal encodings. The encodings of literals are interpreted as elements of $\FF_{2^k}$.
For instance, the clause $C = \xzero \lor \neg y \lor \neg \xone$ is encoded as:
\begin{align*}
     \gamma(C)(X) = (X - \texttt{100}_{\FF})(X - \texttt{001}_{\FF})(X - \texttt{010}_{\FF})
\end{align*}
Such clause encoding allows efficient verification of  resolution and reduction steps via arithmetic over $\FF_{2^k}$. 

Consider verifying the correctness of the clause $C_r$, which is claimed to be the result of applying a universal reduction (\ured) step to the original clause $\xzero \lor y \lor \neg \xone$. Let  $P_{C_r}(X)$ be the polynomial encoding $C_r$;
We can first construct the extended witness containing a residual set $W_{\mathsf{res}} = \{\mathtt{100}\}$, a removal set $W_{\mathsf{rem}} = \{\mathtt{010}\}$, and an existential pivot encoding $w_{\exists} = \mathtt{001}$.  Verification proceeds by checking in ZK:  
\begin{enumerate}
\item Membership of $w_{\exists} \in \lexists$ and inclusion of $W_{\mathsf{rem}}\subseteq \lforall$; 
\item Equivalence that $(X - w_{\exists})\prod_{e \in W_{\mathsf{res}}}(X-e) \equiv \gamma({C_r})(X)$; 
\item For each $\ell \in W_{\mathsf{rem}}$, $\getorder(\ell) > \getorder(w_{\exists})$.
\end{enumerate}
Here, $\getorder(\cdot)$ is an operator that maps a literal encoding to its order in the quantifier prefix, defined as the integer represented by the last two bits of the encoding. We also need to verify the dual interpretation of literal encodings, clause containment of literals, and quantifier consistency. Our approach is detailed in Section~\ref{sec:encoding}.

\smallskip
\noindent
{\bf Proof via winning strategies}. In fact, to show the QBF is unsatisfiable, we can further define the functions $f_{\xone}$ and $f_{\xzero}$ for the universal-quantified variables $\xone$ and $\xzero$ as $f_{\xzero} = {\sf false} \quad f_{\xone}(y) = y$. We then substitute the $\xone$ and $\xzero$ with this function and obtain:
\begin{align}
\psi_H =
(y \lor y) 
\land (\neg y \lor \neg y) 
\land ( y \lor \neg y) 
\land ( \neg y \lor y)
\label{eq:herbrand}
\end{align}
Simplifying the formula yields $y \land \neg y$, which is unsatisfiable. This unsatisfiability of $\Psi_H$ directly implies that the original~\Cref{eq:qbf} is false.
The reason is that, using the universal strategy $f_{\xzero} = {\sf false}$ and $f_{\xone}(y) = y$, the universal player can always force the formula to evaluate to false, regardless of the existential player’s choice of $y$. Moreover, since $\xone$ appears after $y$ in the quantifier prefix, $f_{\xone}$ is allowed to depend on $y$ while $f_{\xzero}$ can not.
By such substitutions, we reduce the falsification of the QBF to checking the unsatisfiability of a purely propositional formula, which can be handled using existing ZKP for unsatisfiability proofs for private formulae~\cite{zkunsat}. Meanwhile, we also need to check the following constraints in ZK: 
\begin{enumerate}
    \item \textit{Dependency correctness:} 
    The strategy must respect the quantifier prefix.  Here, $f_{\xzero}$ must be a constant (as $\xzero$ is the first variable), and $f_{\xone}$ can depend only on existential variables that appear before $\xone$ (in this case, $y$).
    
    \item \textit{Substitution correctness:} $\Psi_H$ must be a result from substituting the universal variables $\xzero$ and $\xone$ with their respective strategy functions in the $\psi$.
\end{enumerate}
In Section~\ref{sec:winnin_strategies}, we describe how efficient verification of dependency correctness and substitution correctness can be achieved in ZK. 

\section{Preliminaries}
\label{sec:preliminaries}

\subsection{Quantified Boolean Formulae}
A \emph{quantified Boolean formula (QBF)} is a propositional formula extended with quantifiers over Boolean variables. A QBF in \emph{prenex conjunctive normal form (PCNF)} has the form:
\[
\Psi = \underbrace{Q_1 \var_1 \; Q_2 \var_2 ; \dots \; Q_n \var_n.}_{\text{ prefix}}\; \underbrace{\psi(\xzero, \dots, x_n)}_{matrix}
\]
where each $Q_i \in \{\forall, \exists\}$ is a quantifier and $\var_i$ is a set of Boolean variables. 
$\psi$ is a Boolean formula in conjunctive normal form (CNF) over variables drawn from $ \var = \bigcup{\var_i}$.
Throughout this work, we assume that all quantified Boolean formulae (QBFs) are given in PCNF. QBF truth evaluation is known to be PSPACE-complete. 

\noindent
{\it Semantics.}
Let $\Phi$ be as above. The truth of $\Phi$ is defined inductively:
\begin{itemize}
    \item[-] If $\Phi$ has no quantifiers, then its value is determined by the truth of the propositional formula $F$.
    \item[-] $\exists x. \Psi$ is true if there exists $b \in \{0,1\}$ such that $\Psi[x \mapsto b]$ is true.
    \item[-] $\forall x. \Psi$ is true if for all $b \in \{0,1\}$, $\Psi[x \mapsto b]$ is true.
\end{itemize}
We use $\varforall$ and $\lforall$ to denote the sets of universally quantified variables and their corresponding literals. Literals are variables or their negations. That is, $\varforall = {x_i \mid x_i \in \var_i,, Q_i = \forall}$ and $\lforall = {x_i, \neg x_i \mid x_i \in \varforall}$. Similarly, we use $\varexists$ and $\lexists$ to denote the sets of existentially quantified variables and literals.

The quantifier prefix $ Q_1\var_1\cdots Q_n\var_n$ induces a partial order over the literal set $\lits$ based on the ordering of quantifier blocks. 
Specifically, for $x_i \in \var_i$ and $x_j \in \var_j$, we write $x_i \prec x_j$ (and similarly $\neg x_i \prec x_j$, $x_i \prec \neg x_j$, etc.) if the block $\var_i$ appears before $\var_j$ in the prefix.

A clause, defined as a disjunction of literals, is naturally represented as a \emph{set} of literals. In this representation, the notation $\ell\in C$ denotes that the literal $\ell$ occurs as a disjunct in clause $C$. The union $C_1 \cup C_2$ corresponds to the clause containing all literals that appear in either $C_1$ or $C_2$. 

\smallskip\noindent 
\textbf{The Q-Resolution proof system (\qres).}
A \emph{Q-resolution proof} operates on the clauses in the CNF matrix of a QBF and derives the \emph{empty clause} using the following rules:

\noindent
\textit{Universal Reduction (\ured).} Let $\Psi = Q.\psi$ be a QBF in PCNF and let $C$ be a clause in $\psi$. Let $y$ be the innermost existential literal in $C$. Any universal literal $x_i \in C$ with $y \prec x_i$ can be removed from $C$ without affecting the truth value of $\Psi$. We say that $C \vdash_{Q, \forall-RED} C_r$ if $C_r$ is obtained by applying \ured{} on $C$ according to $Q$.

\smallskip
\noindent
\textit{Existential-Resolution (\xres).} Let $C_1, C_2$ be clauses in $\psi$. If there exists an existential variable $y \in \var_\exists$ such that $y \in C_1$ and $\neg y \in C_2$, the Q-resolvent of $C_a$ and $C_b$ on pivot $y$ is derived as follows by computing the resolvent $C = (C_a \cup C_b) \setminus \{y, \neg y\}$. If $C$ is a tautology, discard it; otherwise, $C$ is the Q-resolvent.
We say that $C_a, C_b, y \vdash_{Q, \mathsf{xres}} C_r$ if $C_r$ is obtained by applying \xres{} on $C_a, C_b$ pivoting on $y$ according to $Q$.

A Q-resolution proof is a finite sequence of clauses $C_1, C_2, \ldots, C_m$ in which each clause $C_i$ is either an initial clause from the matrix $F$ or is derived from earlier clauses $C_a$ and $C_b$ by first applying universal reduction (\ured) to each and then performing existential resolution (\xres) on the results. If the final clause in the sequence is the empty clause $\bot$ (i.e., $C_m = \bot$), the sequence constitutes a \emph{Q-resolution refutation} of the QBF $\Phi$.

\begin{theorem}[~\cite{BUNING199512}]
\label{thm:qres}
A closed QBF in PCNF is unsatisfiable if and only if there exists a sequence of Q-resolution steps leading to the empty clause.
\end{theorem}

\noindent To prove a QBF to be true, one can employ a dual proof system. This proof system can be implemented in zero-knowledge using techniques almost identical to those used for \qres. We provide details in Appendix~\Cref{sec:cube}.

\paragraph{Winning strategies.}
A \emph{winning strategy} for the universal player, i.e., a refutation strategy, is a set of Boolean functions
\[
\mathcal{H} = \{ g_{x} \mid x \in \varforall \}
\]
such that for some $x \in \varforall$, the function $g_x: \mathrm{Pred}_\exists(x) \to \{0,1\}$ maps the valuations of all variables that precede $x$ in the prefix to a Boolean value. Here, $\mathrm{Pred}_\exists(x) = \{ x_j \mid x_j \prec x \}$. The strategy $\mathcal{H}$ is \emph{winning} for the universal player if substituting each universal variable $x \in \lforall$ with its corresponding \emph{Herbrand function} $g_x$ yields a propositional formula $\Psi_H$ that is unsatisfiable. We refer to $\Psi_H$ as the \emph{Herbrandization} of the QBF $\Psi$.

Dually, a winning strategy for the existential player is a set of Boolean functions $
\mathcal{S} = \{ f_{x} \mid x \in \varexists \}
$
such that for every $x \in \varexists$, the function $f_x : \mathrm{Pred}_\forall(x) \to \{0,1\}$ maps the valuations of all \emph{universally} quantified variables that precede $x$ in the quantifier prefix to a Boolean value. 
The strategy $\mathcal{S}$ is \emph{winning} if substituting each existential variable $x \in \varexists$ with its corresponding \emph{Skolem function} $f_x$ in $\psi$ yields a propositional formula $\Psi_S$ over only the universal variables $\varforall$ that is a tautology.  We refer to $\Psi_S$ as the \emph{Skolemization} of the QBF $\Psi$.

\begin{theorem}[\cite{balabanov2011resolution, niemetz2012resolution}]
    A QBF $\Psi$ is \emph{true} (or satisfiable, when there are free variables) if and only if there exists an existential winning strategy (i.e., Skolem functions). It is \emph{false} if and only if there exists a universal winning strategy (i.e., Herbrand functions).
\end{theorem}

\smallskip\noindent
{\it Example.} We define the Herbrand functions $f_{\xone}$ and  $f_{\xzero}$  for $\xone$ and $\xzero$ as:
$
f_{\xzero} = {\sf false} \quad f_{\xone}(y) = y
$.
We substitute the universal variables with their corresponding Herbrand functions and obtain an unsatisfiable formula. Therefore, the strategy above is a valid Herbrand strategy, and the QBF is \emph{false}.

\subsection{Efficient Zero-Knowledge Proof}
ZKP~\cite{STOC:GolMicRac85,GolMicWig91} allows a prover to convince a verifier that it possesses an input $w$ such that $P(w) = 1$ for some public predicate $P$, while revealing no additional information about $w$. There have been many lines of work in designing practically efficient ZK protocols under different settings and assumptions (eg. ~\cite{STOC:IKOS07,STOC:GolKalRot08,AC:Groth10a,CCS:JawKerOrl13}).

The goal of this work is to build practical ZKPs for QBF evaluations and winning strategies based on existing protocols rather than proposing the construction of a general-purpose ZKP. To this end, 
we extract the ZKP functionalities required for our protocol in Figure~\ref{func:zk}. In particular, we use a special type of ZK protocol commonly referred to as ``commit-and-prove'' ZK~\cite{STOC:CLOS02}, which allows a witness to be committed and later proven over multiple predicates while ensuring consistency of the committed values. Figure~\ref{func:zk} lists the ZK functionality we need for building our protocols. 

\begin{figure}[!t]
   \begin{nffunc}{$\Func[ZK]$}
    \shortsection{Witness:}
On receiving $({\sf Witness}, x)$ from the prover, where $x\in\FF$, store $x$ and
send $[x]$ to each party.

\shortsection{Instance:}
On receiving $({\sf Instance}, x)$ from both parties, where $x\in \FF$, store $x$ and
send $[x]$ to each party. If the inputs sent by the two parties do not match, the functionality aborts.

\shortsection{Circuit relation:}
On receiving $({\sf Relation}, C, [\xone],\ldots,$ $[x_{n-1}])$ from both parties,
where $x_i\in\FF$ and $C\in\FF^n\rightarrow\FF^m$,
compute $y_1,\ldots,$$y_m:=C(\xone,\ldots,x_{n-1})$ and send $\{[y_1],\ldots,[y_m]\}$ to both parties.

\shortsection{Productions-of-multi-variate-polynomial equality:} On receiving $({\sf PoPEqCheck}, X$ 
   $\{[P_i(X)]\}_{i\in[n]},  \{[Q_i(X)]\}_{i\in[m]})$ from both parties, where $[P_i(X)]$ and $[Q_i(X)]$ are multi-variate polynomials over $X$ with their coefficient committed: if $\Pi_iP_i(X)\neq \Pi_i Q_i(X)$, the functionality aborts.

\shortsection{Conversion:}
On receiving $({\sf Conv}, [x], i, \xi)$ from $\Prv$ and $({\sf Conv}, [x], \xi)$ from $\Ver$ where $i\in \N$, store $i$ and send $[i]$ to both parties. If $\xi: \FF\rightarrow \N$ from $\Prv$ and $\Ver$ are different or $\xi(x) \neq i$, the functionality aborts.  

\shortsection{Comparison:}
On receiving $({\diamond}, [i], [j], \lambda)$ from both parties, 
where $\diamond\in\{>, \geq, <, \leq\}$. If $i$ or $j \notin N$ or  $i \diamond j$ does not hold when $\diamond$ is interpreted as the standard integer comparison, the functionality aborts. 

\end{nffunc}
\caption{Functionality for zero-knowledge proofs of circuit satisfiability and integer comparison.}
    \label{func:zk}
\end{figure}

\paragraph{Set argument.} Our protocol reduces statements over literals and clauses to statements about set membership and subset relations. These set-based primitives serve as the foundation for verifying structural properties of formulae. In practice, a variety of ZKPs exist that efficiently implement such set operations, such as~\cite{benarroch2023zero}. We list the functionality of our interests in~\Cref{idfunc:set}.
\begin{figure}[h]
      \begin{nffunc}{$\Func[ZKSet]$}

   \shortsection{Set initialization:}
      On receiving $({\sf Init}, N, [s_1],\ldots,$ $[s_{N}])$ from \Prv and \Ver, where
      $s_i\in\FF$. Store the~${S} =\{s_i\}$ and set $f:={\sf honest}$ and send $[S]$ to each party. 

  \shortsection{Set subset:}
  On receiving $({\sf Subset}, S', [S])$ from \Prv, and $({\sf Subset, [S]})$ from \Ver,
  If ${S'} $ is not a subset of ${S}$, set $f:={\sf cheating}$ and send $[S']$ to each party.  

  \shortsection{Set membership:}
  On receiving $({\sf Mem}, \{[s_1], \cdots, [s_w], [S]\})$ from \Prv and \Ver, where ${s_i} \in\FF$, set $f:={\sf cheating}$  if $s_i \notin {S}$ for some $i$.

\shortsection{Set check:} Upon receiving $({\sf check})$ from $\Ver$ do: If \Prv sends $({\sf cheating})$ then send {\sf cheating} to $\Ver$. If \Prv sends $({\sf continue})$ then send     $f$ to~$\Ver$.
   
     \end{nffunc}
     \caption{Functionality for set operations in ZK.}\vspace{-4mm}
     \label{idfunc:set}
 \end{figure}

\paragraph{Append-only array.} We leverage $\Func[FlexZKArray]$  to store all clauses from both the input formula and the derived proof. $\Func[ZKFlexZKArray]$ is a specialized array structure that avoids the overhead of generic RAM access in ZKP by supporting only two operations: \emph{append} and \emph{read}~\cite{zkunsat}. The protocol assumes that the prover precomputes and appends all clause entries in advance, allowing the verifier to verify only the read operations in ZK. 
In addition, the access functionality is intentionally weakened: the verifier cannot track repeated appends or enforce strict ordering of reads. The formal definition is provided in Appendix~\Cref{idfunc:roram}.

\paragraph{ZKUNSAT~\cite{zkunsat}.}
ZKUNSAT proves the unsatisfiability of a private proposition Boolean formula in CNF by leveraging its resolution proof. 
A resolution proof is a sequence of derived clauses, ending in the empty clause $\bot$, where each clause is either part of the initial formula or derived using the resolution rule. The resolution rule allows deriving a new clause from two clauses that contain a unique pair of complementary literals. Specifically, we say that $C_a, C_b, \ell_p \vdash_{\mathsf{res}} C_r$ 
\begin{align*}
     C_a = \ell_p \lor \ell_1 \lor \cdots \lor \ell_m
    \quad  &C_b  = \neg \ell_p \lor \ell'_1 \lor \cdots \lor \ell'_k \\
     C_r = \ell_1 \lor \cdots &\lor \ell_m \lor \ell'_1 \lor \cdots \lor \ell'_k,
\end{align*}
and $C_r$ is not a tautological clause (i.e., it does not contain both a literal and its negation).

Notice that the derivation $C_a, C_b, \ell_p \vdash_{\mathsf{xres}} C_r$ is the same as the \xres step in \qres, with the exception that it does not enforce the pivot variable to be existentially quantified. We leverage the efficient technique from~\cite{zkunsat} to verify resolution steps in \qres. Below, we outline how~\cite{zkunsat} enables ZKP of correct resolution steps and the conditions under which soundness is guaranteed. 

In ZKUNSAT, each literal $\ell$ is encoded as a field element $\epsilon(\ell) \in \FF$ and committed accordingly. A clause $C$ is represented as a polynomial $\polynomial_C$ whose roots are the encoded literals in $C$. To verify the correctness of resolution steps, {ZKUNSAT} checks a set of algebraic relations over the polynomials $\polynomial_C$, $\polynomial_C'$, and $\polynomial_{C_r}$ corresponding to the two parent clauses and their resolvent. These checks reduce to verifying polynomial identities, which can be implemented using $\Func[ZK].{\tt PoPEqCheck}$. To ensure the soundness, the encoding function $\epsilon(\cdot)$ should be injective and designed such that for every literal $\ell$, and $\epsilon(\ell) + \epsilon(\neg \ell) = \texttt{cst}_{\FF}\footnote{This condition is necessary to enable efficient checking of complementary literals.}$ holds.

for some fixed constant $\texttt{cst}_{\FF} \in \FF$. Under this encoding, satisfaction of the polynomial relations guarantees the soundness of each resolution step in the proof. 
ZKUNSAT also leverages the $\Func[FlexZKArray]$ functionality to store all clauses from both the input formula and the derived proof.

\section{ZKP for \qres Proof Validation}
\label{sec:methodology}
This section presents our protocol for verifying the correctness of a QBF's Q-resolution proof. We begin by describing the data structure used to encode QBFs and \qres proofs arithmetically. We then detail how this structure enables the efficient verification of \ured{} and \xres{} steps within a zero-knowledge proof (ZKP) framework.

\subsection{Encoding QBF}
\label{sec:encoding}
\subsubsection{Ordered literal encoding} The variables in QBF are ordered. Given a QBF with $\Psi = {Q_1 \var_1 \;\dots \; Q_k \var_k.}$ ${\psi(x_1,\dots,x_n)}$ over variables $\var= \cup \var_i$. We assume that the quantifier prefix induces a total order over $\var$.\footnote{We assign each quantifier block $\var_i$ a total order over its variables and extend this to a total order over the entire set $\var$, such that the global ordering is consistent with the quantifier prefix.} We define the encoding of a literal $\ell$ as the concatenation of three \emph{binary strings}:
\begin{align}\label{eq:encoding}
    \epsilon(\ell) = \order(\ell)_{\bf B}  \|\ \textsf{sign}(\ell)
\end{align}
where $\textsf{sign}(\ell) \in \{0,1\}$ is a 1-bit flag indicating the sign of the literal: $\textsf{sign}(\ell) = 0$ if $\ell = \neg x$, and $\textsf{sign}(\ell) = 1$ if $\ell = x$ for some $x \in \var$. The function $\order(\ell) = \order_x $ if $\ell = x$ or $\neg x$, and $ \order_x$ is the order of $x$ in $\var$. We use $\order(\ell)_{\bf B}$ to denote the binary presentation of the order as an integer. 

We adopt this encoding to enable efficient extraction of the information required for verifying both \ured{} and \xres{} steps. 
The length of encoding can be bounded by $1 + \log {\var}$. In addition, the encoding is injective when the string length is fixed, enabling each literal to be uniquely mapped to an element in $\FF$.

Each encoding $\epsilon(\ell) \in \{0,1\}^{k}$ is interpreted both as a unique element in a finite field and as a natural number via the interpretation functions ${\sf Itp}{\FF}$ and ${\sf Itp}{\N}$, respectively:
\begin{align}
{\sf Itp}_{\FF}: &~\{0,1\}^k \rightarrow \FF  \text{ s.t. }  {\sf Itp}_{\FF}(\epsilon(\ell)) + {\sf Itp}_{\FF}(\epsilon(\neg \ell)) = 1_{\FF}\label{eq:itpf} \\
{\sf Itp}_{\N}: &~\{0,1\}^k \rightarrow \N  \text{ s.t. }   {\sf Itp}_{\N}(\epsilon(\ell)) = {\sf order}(\ell)\label{eq:itpn}
\end{align}
The finite field $\FF$ is chosen such that $|\FF| \geq |\lits|$, and the function ${\sf Itp}_{\FF}$ should also be injective.

The composed function ${\sf Itp}_{\FF} \circ \epsilon$ provides an injective encoding that satisfies the algebraic constraints needed to verify resolution steps in zero knowledge using the approach introduced in~\cite{zkunsat}. Meanwhile, ${\sf Itp}_{\N} \circ \epsilon$ allows for recovering the relative quantifier order of literals, which is necessary to verify the correctness of universal reduction (\ured).

The remaining task is to verify the correctness of conversions between elements in the finite field $\FF$ and their corresponding bounded integer representations in $\mathbb{N}$ in ZK. We can leverage the existence of interpreters of any given \FF, as we showed in Figure~\ref{prot:zkliteral}. In our implementation, we achieve this by adopting the approach introduced in~\cite{weng2021mystique, weng2021wolverine, baum2021appenzeller}, which provides an efficient approach for checking consistency between $\FF$ and bounded integers in $\mathbb{N}$ in ZK when $\FF$ is instantiated as $\FF_{2^k}$. 

\noindent
\emph{ZK operations on ordered literals.} In~\Cref{func:zkliteral}, we specify the zero-knowledge operations over literals under a total order induced by the quantifier prefix of the input QBF, formalized in the functionality $\Func[OrdLiteral]$.

During initialization, both parties provide an ordering function $\order: \lits \rightarrow \mathbb{N}_{2^{\ordlength}}$ and bit-length parameter $\ordlength$. The functionality ensures that the parties agree on an ordering. 

To commit a literal, prover inputs its encoded value, and both prover and verifier receive a committed version $[\ell]$ via \Func[ZK]. To retrieve the quantifier-induced order of a literal, prover and verifier invoke the \texttt{Order} interface, which checks that the provided rank $i$ matches $\order(\ell)$. The functionality then returns $[i]$ as a committed integer. These operations ensure that ordering checks over literals, necessary for verifying quantifier-respecting dependencies in Herbrand and Skolem strategies, can be performed in ZK.

\begin{figure}[!t]
   \begin{nffunc}{$\Func[OrdLiteral]$}
\shortsection{Parameter:} On receiving $({\sf Init}, \order, \ordlength)$ from \Prv and \Ver where $\order: \lits \rightarrow \N_{2^\ordlength}$, abort if they are not identical; otherwise store $(\order, \ordlength)$.

\shortsection{Input:}
   On receiving $({\sf Input}, \ell)$ from \Prv and $({\sf Input})$ from verifier where $\ell \in\lits$, if $\ell$ from two parties differ, the functionality aborts;  Otherwise store $\ell$ and send $[\ell]$ to both parties.

\shortsection{Order:}
   On receiving $({\sf Order}, ([\ell], i))$ from \Prv and $({\sf Order, [\ell]})$ from verifier. If $\order(\ell) \neq i$, the functionality aborts. Otherwise, the functionality sends $[i]$ to both parties.
   \end{nffunc}
    \caption{Functionality for ZK operations on ordered literals.}
     \label{func:zkliteral}
     \vspace{-3mm}
\end{figure}

\begin{figure}[!t]
      \begin{nfprot}{\Prot[OrdLiteral]}
  \shortsection{Parameter:} Given the function $\order$, both $\Prv$ and $\Ver$ compute and agree on the functions ${\sf Itp}_{\mathbb{F}}$, and ${\sf Itp}_{\mathbb{N}}$ that satisfy Equation~\ref{eq:itpf} and~\ref{eq:itpn}.

\shortsection{Input:}
\Prv computes the encoding $\epsilon(\ell)$ according to Equation~\ref{eq:encoding}. Then \Prv and \Ver authenticate ${\sf Itp}_\FF(\epsilon(\ell))$ using \Func[ZK] and obtain $[{\sf Itp}_\FF(\epsilon(\ell))]$. The two parties then output $[\ell] = [{\sf Itp}_\FF(\epsilon(\ell))]$.

\shortsection{Order:}
Set $\xi={\sf Itp}_\N$$\cdot{\sf Itp}^{-1}_\FF$. \Prv sends $({\sf NInterpret}$, $[\ell], i, \xi)$ to \Func[ZK], while \Ver sends $({\sf NInterpret},$$ [\ell], \xi)$. They set the output as the returned value $[i]$.
   
\end{nfprot}
    \caption{Protocols for ZK operations on ordered literals.}
     \label{prot:zkliteral}
          \vspace{-6mm}

\end{figure}

\subsubsection{Clause encoding} We adopt clause encoding from~\cite{zkunsat} for efficient verification of \xres{} and \ured{} in ZK, which we elaborate in~\ref{sec:preliminaries}. The clause is encoded as a polynomial over $\FF$ rooted at the encodings of literals in the clauses, when the encodings are interpreted as elements in $\FF$. Formally, for $C= \ell_1 \lor \cdots \lor \ell_d$ 
\begin{align*}
\gamma(C)(X) =(X - {\sf Itp}_\FF(\epsilon(\ell_0))) \cdots (X - {\sf Itp}_\FF(\epsilon(\ell_d))).
\end{align*}

\noindent
\emph{ZK operations on polynomial-encoded clauses.}
We define a functionality $\Func[Clause]$ that supports authenticated operations over clauses in ZK, described in Figure~\ref{func:zkclause}. The protocols for implementing $\Func[Clause]$ are explained in Appendix~\Cref{protocol:zkclause}. The prover initializes a clause by inputting a list of literal encodings, whose size is bounded by a public parameter $w$, and both parties receive a shared commitment $[C]$. Given this commitment, the parties can retrieve literals in the clauses using \texttt{Retrieval}, which ensures clause consistency by checking the inclusion of declared literals. 

The functionality also provides a \texttt{Equal} operation that verifies in ZK whether a clause is logically equivalent to the disjunction of a set of committed subclauses, requiring no duplicate literal inclusion across clauses. 

For resolution steps, $\Func[Clause]$ offers a \texttt{Res} operation that checks whether a resolvent $[C_r]$ is validly derived from two input clauses under the non-tautological resolution rule. Finally, the \texttt{IsFalse} interface allows checking whether a clause is the empty clause $\bot$, signaling refutation in QBF proofs. All operations are designed to be sound and privacy-preserving within the zero-knowledge framework.

\begin{figure}[!t]
   \begin{nffunc}{$\Func[Clause]$}
\shortsection{Input:}
   On receiving $({\sf Input}, \ell_0, \cdots, \ell_{k-1}, w)$ from \Prv and $({\sf Input}, w)$ from verifier where $\ell_i \in \lits$, the functionality check that $k\leq w$ and abort if it does not hold. Otherwise store $C = \ell_0\lor  \cdots \lor \ell_{k-1}$, and send $[C]$ to each party. 

   \shortsection{Literal retrieval:}   On receiving $({\sf Retrieval}, \{\ell_1, \cdots, \ell_k\}, [C])$ from $\Prv$ and $({\sf Retrieval}, [C])$ from $\Ver$, check if $C = \{\ell_1 \lor \cdots\lor \ell_w\}$; if not the functionality aborts. Otherwise, send $[\ell_1],\cdots,[\ell_k]$ to each party.

   \shortsection{Equal:} Upon receiving $({\sf Equal}, [C], \{[C_i]\})$ from both parties, the functionality first checks whether any two clauses $C_i$ and $C_j$ (for $i \neq j$) have overlapping literals. If such overlap exists, the functionality aborts. Otherwise, it verifies whether $C = \bigvee_i C_i$; if this condition is not satisfied, the functionality aborts.
   
    \shortsection{Res:}  On receiving $({\sf res}, [C_0], [C_1], [\ell_p], [C_r])$ from both parties, check if $\{C_0, C_1\}, \ell_p \vdash_{\tt res} C_r$; if not the functionality aborts.

     \shortsection{IsFalse:}  On receiving $({\sf IsFalse},  [C])$ from both parties, check if $C = \bot$; if not, the functionality aborts.

   \end{nffunc}
    \caption{Functionality for ZK operations on clauses.}\vspace{-4mm}
     \label{func:zkclause}
\end{figure}

  \begin{figure}[!t]
      \begin{nfprot}{\Prot[Clause]}
      \shortsection{Parameters:} A set $\lits$ of all possible literals and a finite field $\FF$. An integer $w$ and a set of clauses $\mathbf{C}_w$ that contains all clauses no more than $w$ literals of $\lits$. $\epsilon: \lits \rightarrow \{0,1\}^k$ and ${\sf Itp}_\FF$ are injective.
      
    \shortsection{Input~\cite{zkunsat}:} 
    \begin{enumerate}[leftmargin = 4mm]
    \item $\Prv$ holds a clauses $C = \ell_0 \vee \cdots \vee \ell_{k-1} \in \mathbf{C}_w$, defines $ \gamma(C)(X) =(X - {\sf Itp}_\FF(\epsilon(\ell_0))) \cdots (X - {\sf Itp}_\FF(\epsilon(\ell_d)))$ and
    locally computes $c_0,\ldots,c_w$ such that $\gamma(C)(X) = \sum_{i\in [0, w]} c_iX^i$.
    \item For each $i\in [0, w]$, two parties use \Func[ZK] to get $[c_i]$.
    Two parties output $[\gamma(C)] = \{[c_i]\}_{i\in[0,w]}$
    \end{enumerate}
    
\shortsection{Equal:}  Both parties send $({\sf PoPEqCheck}, X, [\gamma(C)(X)],$ $\{[\gamma(C_i)(X)]\} )$ to $\Func[ZK]$. 

  \shortsection{Literal retrieval:} 
 \begin{enumerate}[leftmargin = 4mm]
 \item $\Prv$ locally computes the encodings $\epsilon(\ell_i)$ and authenticate $\rho_i(X) = X-\epsilon(\ell_i)$ using $\Func[ZK]$. AS a result, two parties get $[\rho_i(X)]$.
 \item Both parties send $({\sf PoPEqCheck}, X, \{[\rho_i(X)]\}, \{\gamma(C)\}(X))$ to $\Func[ZK]$.
 \end{enumerate}

  \shortsection{Res~\cite{zkunsat}:} Details are provided in Appendix~\Cref{protocol:clauseres}.

    \shortsection{IsFalse~\cite{zkunsat}:} Both send $({\sf PoPEqCheck},X, [\gamma(C)(X)]$, $[1])$. 
      \end{nfprot} 
     \caption{Our protocol to instantiate \Func[Clause].}
     \label{protocol:zkclause}
 \end{figure}

 \vspace{12pt}

\subsubsection{Quantifier encoding} We leverage two \emph{public} sets $\lforall = \{x,\neg x | x \in \varforall\}$ and $\lexists = \{x,\neg x | x \in \varexists\}$ for keeping quantifier information with the order of the variables encoded by the literal encoding. Notice these two sets are public information as the QBF is public. The verifier can directly check if these two sets are well-formed: for any $\ell\in\lforall$ (or $\lexists$), $\neg\ell$ is also in $\lforall$ ($\lexists$). We list the functionality of $\Func[Quantifier]$ in Figure~\ref{func:quantifier}. This functionality can be directly realized using $\Func[ZKSet]$.

\begin{figure}[!t]
   \begin{nffunc}{$\Func[Quantifier]$}
\shortsection{Initialization:} Upon receiving $({\sf init}, \lexists, \lforall)$ from both $\Prv$ and $\Ver$, where $\lexists$ and $\lforall$ are sets of literals, the functionality stores both lists. Abort if the inputs are not consistent across the two parties. Otherwise send $[\lforall]$ and $[\lexists]$ to each party. Set $f:={\sf honest}$ and ignore subsequent initialization calls.

\shortsection{Check:} On receiving  $({\sf check}, [\ell], \Box)$ from both \Prv and \Ver, where $\Box \in \{\forall, \exists\}$. Set $f:={\sf cheating}$ if $\Box$ from parties are not consistent, or $\ell \notin \mathcal{L}_{\Box}$.
   \end{nffunc}
    \caption{Functionality for ZK operations on quantifier.}\vspace{-4mm}
     \label{func:quantifier}
\end{figure}

\subsection{ZKP of QBF Evaluation via \qres Proofs}
A Q-resolution (\qres) proof consists of a sequence of inference steps, each applying the \qres rule. Every \qres step takes two supporting clauses as input and performs an existential resolution (\xres{}) followed by a universal reduction (\ured{}) on the resulting resolvent. To ensure correctness, the verifier must check that both supporting clauses are either part of the input QBF's matrix or have been derived in earlier \qres steps.
The latter condition can be handled using append-only data structures in ZK, following the approach in~\cite{zkunsat}. In this section, we first focus on how to verify the correctness of  \ured{}(~\Cref{sec:ured}) and \xres{}(~\Cref{sec:xres}) steps in zero knowledge using the literal encoding described in Section~\ref{sec:encoding}. 

\subsubsection{\bf Existential resolution}\label{sec:xres}  
Given two committed clauses $C_a$ and $C_b$, $\{C_a, C_b\}, \ell_p$$\vdash_{\sf Q,xres}$ $ C_r$ if and only if the following conditions hold:

\noindent
\begin{minipage}[b]{.3\linewidth}
\begin{equation}
\ell_p \in \lexists \label{eq:xres_quantifier}
\end{equation}
\end{minipage}
\hfill and
\begin{minipage}[b]{.5\linewidth}
\begin{equation}
 C_a, C_b, \ell_p \vdash_{\sf res} C_r \label{eq:xres_resolution}
 \end{equation}
\end{minipage}

\noindent
Here, $\ell_p$ denotes the pivot literal, and $\vdash_{\sf res}$ denotes the non-tautological resolution rule applied over complementary literals $\ell_p \in C_a$ and $\neg \ell_p \in C_b$. Constraint~\eqref{eq:xres_quantifier} ensures that the pivot literal is existentially quantified, as required by the semantics of \qres.
The verification of \xres{} steps follows the same approach as in~\cite{zkunsat}, with the additional requirement of checking that the quantifier type of the pivot literal is existential. This ensures that resolution is applied only over variables whose assignments are controlled by the existential quantifier, preserving the soundness of the proof under the QBF semantics.

\subsubsection{Universal reduction}
\label{sec:ured}Given two committed clauses $C$ and $C_r$, $C_r$ is a valid universal reduction of $C$ if and only if the following hold:
\begin{align}
    \exists w_\ell, \quad W_{\sf res} &= \ell_1 \lor \cdots \lor \ell_k,\quad 
    W_{\sf rem} = \ell'_1 \lor \cdots \lor \ell'_d \quad \text{s.t.} \nonumber \\
     \order(\ell_i) &<\order(w_\ell), \quad \forall i \in [k] \label{eq:order_res} \\
      \order(\ell'_i) &>\order(w_\ell), \quad \forall i \in [d]\label{eq:order_rem}  \\
    \ell'_i &\in \lforall, \quad w_\ell \in \lexists \label{eq:quantifier} \\
    C = W_{\sf res} &\cup W_{\sf rem} \cup \{w_\ell\}, \quad    C_r = W_{\sf res} \cup \{w_\ell\} \label{eq:equivalence}
\end{align}

Here, $W_{\sf rem}$ denotes the set of universally quantified literals that are removed during the \ured{} step, while $W_{\sf res}$ contains the remaining literals preserved in the reduced clause. The literal $w_\ell$ is the existentially quantified literal in $C$ with the highest order according to the quantifier prefix. Prover can prepare  $w_\ell, \{w_\ell\}, W_{\sf res}$ and $W_{\sf rem}$ and commit them via  $\Func[ZKLiteral]$ and $\Func[ZKClause]$. 

In a universal reduction step, a set of universally quantified literals is removed from clause $C$ to obtain a smaller clause $C_r$. This operation is sound only if the removed literals occur at the innermost positions in the quantifier prefix. Constraints~\ref{eq:order_rem} and~\ref{eq:quantifier} together ensure this condition holds: the removed literals must be universally quantified and must appear after all other literals in $C$ with respect to the quantifier order.
The soundness of Q-resolution (\qres) requires that all derived clauses be universally reduced. This means that any universal literals appearing after the innermost existential literal in a clause must be eliminated. Constraints~\ref{eq:order_res} and~\ref{eq:quantifier} enforce this condition by ensuring that only the subset of universal literals with higher quantifier levels than the last existential literal is removed.
Finally, constraint~\ref{eq:equivalence} guarantees that the decomposition of clause $C$ into the retained literals ($W_{\sf res}$), the removed literals ($W_{\sf rem}$), and the retained existential literal ($w_\ell$) is correct, and that the resulting clause $C_r$ is computed accordingly. Together, these constraints ensure the correctness and soundness of each universal reduction step in the \qres proof.

\begin{theorem}\label{thm:qrescorrectness}
Let $\Psi$ be a QBF in PCNF, where all clauses are already universally reduced. Let its \qres proof consist of a sequence of derivation steps, each of the form $(C_a, C_b) \vdash C_r$, where $C_r$ is derived from premises $C_a$ and $C_b$. The proof is sound if the following conditions hold for every step:
\begin{enumerate}
    \item The clauses $C_a$ and $C_b$ are either initial clauses from the matrix of $\Psi$ or were derived in previous steps of the proof.
    \item There exists a pivot literal $\ell_p$ and an intermediate clause $\bar{C}_r$ such that:
    \begin{itemize}
        \item $C_a$, $C_b$, $\ell_p$, and $\bar{C}_r$ satisfy the resolution constraints in~\Cref{eq:xres_quantifier,eq:xres_resolution}, and
        \item $\bar{C}_r$ and $C_r$ satisfy the universal reduction and structural consistency conditions in~\Cref{eq:order_res,eq:order_rem,eq:quantifier,eq:equivalence}.
    \end{itemize}
\end{enumerate}
Under these conditions, the final clause $C_r = \bot$ constitutes a valid \qres proof for refuting $\Psi$.
\end{theorem}
This theorem follows directly from~\Cref{thm:qres} and definitions of \ured and \xres.

\subsection{\qres Validation in ZK}
We present our protocol to verify the correctness of a \qres proof in ZK, detailed in Figure~\ref{protocol:main}. First, both parties obtain clause commitments $[C_i]$ for all clauses in the input formula using $\Func[Clause]$. We adopt the weakened random array access approach proposed in~\cite{zkunsat} to store both the clauses from the QBF matrix and the derived clauses obtained through consecutive applications of \xres{} and \ured{} steps.  In particular,  prover extracts each step of the \qres proof locally, including all derived clauses, and both parties authenticate all clauses (input and derived) and store them in an array initialized using $\Func[FlexZKArray]$.

\begin{figure*}[t]
  \begin{nfprot}{CheckProof}
    \shortsection{Inputs:} Both parties are given a QBF of the form $Q_1 \var_1 Q_2 \var_2 \cdots Q_k \var_k \psi$, where $\psi = C_1 \lor \cdots \lor C_\flength$.  
    The prover (\Prv) holds a \qres refutation encoded as a sequence of tuples $(k_1, l_1), \ldots, (k_R, l_R)$.  
    Both parties know the proof length $R$, proof width $w$, and deduction degree $d$.
    
    \shortsection{Protocol:}
    \begin{enumerate}
      \item \Prv and \Ver obtain $[C_i]$ for $i \in [1, \flength]$ using $\Func[OrdLiteral]$ with the order induced by the quantifier prefix $Q_1 \var_1 \cdots Q_k \var_k$, and $\Func[Clause]$.  
            They also compute $[\lexists]$ and $[\lforall]$ using $\Func[Quantifier]$.
      
      \item \Prv locally derives each clause $C_{\flength + i}$ from the \qres proof and commits them using $\Func[Clause]$.
      
      \item Both parties send $({\sf Init}, \flength + R, [C_1], \ldots, [C_{\flength+R}])$ to $\Func[FlexZKArray]$.
      
      \item For each $i \in [1, R]$, the parties perform the following steps:
        \begin{enumerate}
          \item \Prv looks up the tuple $(k_i, l_i)$ such that $C_{k_i}$ and $C_{l_i}$ resolve on pivot $\ell_i$ to yield $\bar{C}_i$, and $\bar{C}_i$ reduces via \ured to $C_i$.
                The parties compute $[\ell_i]$ and $[\bar{C}_i]$ using $\Func[OrdLiteral]$ and $\Func[Clause]$.

          \item \textbf{Fetching premises:}  
                \Prv sends $({\sf Read}, l_i, C_{l_i}, i)$ to $\Func[FlexZKArray]$,  
                while \Ver sends $({\sf Read}, i)$; both obtain $[C_{l_i}]$.  
                They similarly retrieve $[C_{k_i}]$ and $[C_i]$.
          
          \item \textbf{Checking \xres:}  
                \Prv finds the corresponding pivot literal $\ell_{p_{i}}$ and authenticates it. \Prv and \Ver send $({\sf res}, [C_{l_i}], [C_{k_i}],  [\ell_{p_{i}}], [\bar{C}_i],)$ to $\Func[Clause]$,  
                and $({\sf Check}, [\ell_{p_{i}}], \forall)$ to $\Func[Quantifier]$.
          
          \item \textbf{Checking \ured:}
            \begin{enumerate}
              \item \textbf{Preparing extended witness:} \Prv prepares and commits $W_{\sf res}$, $W_{\sf rem}$, and $w_\ell$ based on Equations~\ref{eq:order_rem}, \ref{eq:order_res}, \ref{eq:quantifier}, and \ref{eq:equivalence}, corresponding to the universal reduction step $\bar{C}i \vdash_{Q, \forall-RED} C_i$.  
            \Prv sends $({\sf Retrieval}, \{\ell_1, \ldots, \ell_k\}, [W_{\sf res}])$ and \Ver sends $({\sf Retrieval}, [W_{\sf res}])$ to \Func[ZKClause], both obtaining $\{[\ell_1], \ldots, [\ell_k]\}$. The same for $W_{\sf rem}$ and obtain $\{[\ell'_1], \ldots, [\ell'_d]\}$. \Prv and \Ver send $({\sf order}, [w_\ell], {\sf ord}_w)$ and $({\sf order}, [w_\ell])$ to $\Func[OrdLiteral]$ to obtain $[{\sf ord}_w]$.

              \item \textbf{Checking~\Cref{eq:order_res}:} 
              For each $[\ell_i] \in \{[\ell_1], \ldots, [\ell_k]\}$, \Prv and \Ver send $({\sf order}, [\ell_i], {\sf ord}_i)$ and $({\sf order}, [\ell_i])$ to $\Func[OrdLiteral]$,  
                    then check $(>, [{\sf ord}_i], [{\sf ord}_w])$ via $\Func[ZK]$. 
              
              \item 
              \textbf{Checking~\Cref{eq:order_rem}:}
              For each $\ell'_i \in \{[\ell'_1], \ldots, [\ell'_d]\}$, \Prv and \Ver send  
                    $({\sf order}, [\ell'_i], {\sf ord}'_i)$ and $({\sf order}, [\ell'_i])$ to $\Func[OrdLiteral]$, then check  
                    $(<, [{\sf ord}'_i], [{\sf ord}_w])$ via $\Func[ZK]$.
              
              \item               \textbf{Checking~\Cref{eq:quantifier}:}
\Prv and \Ver  send $({\sf Check}, [w_\ell], \exists)$ to $\Func[Quantifier]$. For each $\ell'_i \in \{[\ell'_1], \ldots, [\ell'_d]\}$, \Prv and \Ver send $({\sf Check}, [\ell'_i], \forall)$ to $\Func[Quantifier]$

              \item \textbf{Checking~\Cref{eq:equivalence}:} 
              \Prv and \Ver obtain $[C_\ell]$ for $C_\ell = \{w_\ell\}$ using $\Func[Clause]$, then invoke  
                    $({\sf Equal}, [C_r], \{[W_{\sf res}], [C_\ell]\})$ and  
                    $({\sf Equal}, [C], \{[W_{\sf res}], [C_\ell], [W_{\sf rem}]\})$ via $\Func[Clause]$.
            \end{enumerate}
        \end{enumerate}

      \item After all $R$ iterations, the parties use $\Func[Clause]$ to verify that $[C_R] = \bot$; if not, \Ver aborts.

      \item The parties send $({\sf check})$ to both $\Func[FlexZKArray]$ and $\Func[ZKQuantifier]$; if either returns  $({\sf cheating})$ , \Ver aborts.
    \end{enumerate}
  \end{nfprot}
  \caption{Protocol for checking a \qres proof in zero knowledge.}\vspace{-3mm}
  \label{protocol:main}
\end{figure*}

Let the QBF be $Q_1 x_1, Q_2 x_2, \cdots, Q_k x_k.\psi$, where $\psi$ is a CNF formula expressed as $C_1 \land \cdots \land C_f$. We assume that each clause $C_i$ has already been universally reduced. 
The \qres proof is given as a sequence of $R$ resolution steps, each identified by a pair of clause indices $(k_i, l_i)$ indicating the premises used in the $i$-th step. The protocol requires public knowledge of the proof length $R$, the resolution proof width $w$ (the maximum number of literals contained in any clause in the \qres proof), and the deduction degree $d$ (the maximum number of literals removed in any \ured{} step).

The protocol begins by having both parties initialize the necessary encodings. Using the quantifier prefix, they compute a total order over variables and generate structured representations of the initial clauses $C_1, \ldots, C_f$ via functionalities $\Func[OrdLiteral]$ and $\Func[Clause]$. Additionally, quantifier sets $[\lexists]$ and $[\lforall]$ are obtained using $\Func[Quantifier]$, capturing which variables are existentially or universally quantified.
The prover then constructs and commits to each derived clause $C_{f+i}$ (for $i = 1$ to $R$) using $\Func[Clause]$. All clauses, both initial and derived, are appended in a commitment array managed by $\Func[FlexZKArray]$, which supports secure and private clause retrieval for future steps in the protocol.

The core of the protocol proceeds iteratively over each proof step $i \in [1, R]$. In each iteration, the prover identifies the pair $(k_i, l_i)$ and the pivot literal $\ell_i$ used to resolve $C_{k_i}$ and $C_{l_i}$ into an intermediate clause $\bar{C}i$, which is then reduced by universal reduction to derive $C_i$. The necessary clauses $[C_{k_i}]$ and $[C_{l_i}]$. They verify that the pivot $\ell_{p_i}$ is universally quantified and that $\bar{C}i$ is a correct result of resolving $C{k_i}$ and $C_{l_i}$ on $\ell_{p_i}$  using \xres, via the $\Func[Clause]$ and $\Func[Quantifier]$ functionalities.

Next, the protocol verifies the correctness of the universal reduction step $\bar{C}_i \vdash_{Q, \forall-RED} C_i$. To do so, the prover prepares three components: $[W_{\sf res}]$ (literals retained), $[W_{\sf rem}]$ (literals removed), and $[w_\ell]$ (the pivot's witness variable). The prover and verifier use $\Func[OrdLiteral]$ to check that each retained literal precedes $[w_\ell]$ in the quantifier order, and each removed literal follows $w_\ell$. They also verify that $[w_\ell]$ is existential and that all removed literals are universal, using $\Func[Quantifier]$. Finally, \Ver verifies through $\Func[Clause]$ that $[C_i]$ is correctly derived using $[W_{\sf res}]$, $[W_{\sf rem}]$ and $[w_\ell]$. 

Completing all $R$ proof steps, the verifier checks that the final clause $C_{f+R}$ is the empty clause. The protocol ends with both parties sending check calls to ensure that $\Func[FlexZKArray]$ and $\Func[ZKQuantifier]$ do not throw {\sf cheating}. 
\begin{theorem}
The protocol in Figure~\ref{protocol:main} is a zero-knowledge proof of knowledge for falsifying QBFs whose clauses are already universally reduced.
\end{theorem}

Our focus is on enabling efficient use of ZKP for \qres proof validation, rather than proposing a new generic ZK protocol. The resulting protocol satisfies completeness and zero-knowledge in a direct manner. In the case of a corrupted verifier, a simulator can extract the \qres{} proof by extracting clause indices from \Func[FlexZKArray]. Verifying soundness then reduces to ensuring that the extracted proof is valid. This is guaranteed by the consistency of \Func[FlexZKArray], which returns the same clause on repeated reads. As long as each round of interaction for checking \xres{} and \ured{} steps, that includes {\sf res}, {\sf equal}, and {\sf retrieval}, executes without causing \Func[Clause] to abort, and the final {\sf check} step does not trigger {\sf cheating} from either \Func[OrdLiteral] or \Func[Quantifier], the proof is valid according to~\Cref{thm:qrescorrectness}.

We present our ZK protocol for validation of \qres proofs for true QBFs in Appendix~\Cref{sec:cube}.

\section{ZKPs of Winning Strategies}
\label{sec:winnin_strategies}
In this section, we present our protocol for verifying the correctness of a winning strategy as a valid witness for a quantified Boolean formula (QBF). 
We begin by describing how to ensure that the winning strategies are well-formed. Next, we present our approach for verifying the correctness of substituting quantified variables with their corresponding Herbrand or Skolem functions. Finally, we verify the unsatisfiability of the resulting propositional formula; this step follows the technique from~\cite{zkunsat} and is therefore omitted.

\noindent
\subsection{Herbrand Function Well-Formedness.}
A Herbrand function assigns values to a subset of universally quantified variables by expressing each as a Boolean function over variables that precede them in the quantifier prefix. That is, a set of propositional Boolean functions $\mathcal{H} = \{ f_{x}:\mathrm{Pred}_\forall(x) \to \{0,1\} \mid x \in \varforall \}$. We represent $\mathcal{H}$ as a list of tuples:
\[
H = \left\{\, \left(x_i,\; \ell^a_i,\; \ell^b_i\right) \;\middle|\; 0 < i \leq h \,\right\}
\]
where each tuple corresponds to a logical constraint of the form $x_i \iff \ell^a_i \land \ell^b_i$. 
Here, the output variable $x_i$ is either a universally quantified variable or an auxiliary variable used to construct the Herbrand function. Formally, we require $x_i \in \varforall \cup \varaux$, where $\varaux$ denotes a set of auxiliary variables disjoint from the original variable set $\var$. The input literals $\ell^a_i$ and $\ell^b_i$ are drawn from the set $\lits \cup \laux$, where $\lits$ is the set of literals over $\var$, and $\laux = \{ x, \neg x \mid x \in \varaux \}$ is the set of literals over auxiliary variables. 
Notice $H$ can be committed as a list of literal tuples via $\Func[ZKLiteral]$.

We define the \emph{strategy dependency relation} $\prec_{H}$ over $L\cup \laux$ induced by a  falsification strategy $H$ as:
\vspace{-1mm}
\begin{align*}
\prec_{H} := \big\{\, (\ell, \ell') \;\big|\;  (x_i,\, \ell^a_i,\, \ell^b_i) &\in H, \\ 
\ell \in \{\ell^a_i,\, &\neg \ell^a_i,\, \ell^b_i,\, \neg \ell^b_i\}, 
 \ell' \in \{x_i,\, \neg x_i\} \,\big\}.
\end{align*}
\noindent
The relation $\prec_H$ captures direct support dependencies between literals in the Herbrand representation: $\ell'\prec_{H}\ell$ if $\ell$ appears in the support of the variable defining $\ell'$. We extend this to its transitive closure when needed to reason about indirect dependencies.

\vspace{0.5em}
\noindent
The list $H$ corresponds to a \emph{well-formed} Herbrand function w.r.t. a given QBF if the following conditions hold:
\begin{itemize}
    \item[\text{(1)}] \textit{Uniqueness:} Each output variable is universally quantified and defined exactly once. That is, $x_i \neq x_j$ for all $i \neq j$, and $x_i \in \varforall \cup \varaux$.
    
    \item[\text{(2)}] \textit{Acyclicity:} The dependency relation $\prec_H$ induced by $H$ must form a partial order over $\lits \cup \laux$; In other words, the dependency induced by $H$ must be acyclic.
    
    \item[\text{(3)}] \textit{Prefix-consistency:} The order $\prec_H$ is consistent with the quantifier-induced order $\prec$ over $\lits$. That is, for any pair $\ell \prec_H \ell'$, it holds that $\ell \prec \ell'$ under the quantifier prefix of the QBF.
\end{itemize}

These conditions ensure that each variable is uniquely defined, that no circular dependencies are introduced across definitions, and that the Herbrand function respects the quantifier structure of the QBF. We describe how each condition can be efficiently verified:

\begin{itemize}
    \item[\text{(1)}] \textit{Uniqueness:} We verify that $\{x_i\} \subseteq \varforall \cup \varaux$ and contains no duplicates. Notice the set $\varforall \cup \varaux$ is public and contains no duplicate elements. This check ensures that no output variable in $H$ is assigned more than once. 

    \item[\text{(2)}] \textit{Acyclicity:} Assuming $H$ is indexed in topological order w.l.g.. We can check acyclicity by checking:
    \[
    \ell^a_i,\, \ell^b_i \in \lexists \cup \{x_j, \neg x_j \mid j < i\}
    \]
   for $(x_i, \ell^a_i, \ell^b_i)$. This condition ensures that dependencies only refer to previously defined variables, eliminating cycles. We implement this through \Func[FlexZKArray].  

    \item[\text{(3)}] \textit{Prefix-consistency:} We define a function $\dep: \{x_i\} \rightarrow \mathbb{N}$ that computes the dependency level of each output as $
    \dep(x_i) = \max(\order(\ell^a_i), \order(\ell^b_i))$,
and $\order(\ell)$ is defined according to the  order over $\lits$ induced by the quantifier prefix, as follows:
    \[
    \order(\ell) =
    \begin{cases}
        \getorder(\ell) & \text{if } \ell \in \lits, \\
        \dep(x_j)     & \text{if } \ell = x_j \text{ or } \neg x_j \text{ for  auxiliary } x_j.
    \end{cases}
    \]
    The prefix-consistency can be verified by checking in ZK:
    \[
    \dep(x_i) \leq \getorder(x_i) \quad \text{holds for all } x_i.
    \]
   
\end{itemize}

\begin{theorem}[Herbrand Function Well-Formedness]\label{thm:herbrand_condition}
Let $\Psi = Q.\phi$ be a closed QBF with prefix $Q = Q_1 \var_1 \dots Q_k \var_n$, and let $H = \{(x_i, \ell^a_i, \ell^b_i)\}_{i=1}^h$ be a Herbrand strategy expressed as a sequence of definitions. Assume $H$ is indexed in topological order. Then  $H$ is a well-formed Herbrand function for $\Psi$ if it satisfies all of the validation checks described above.
\end{theorem}

We put the proof in  Appendix~\Cref{prf:herbrand}. 

\subsection{Variable Substitution} 
We now describe how to verify the correctness of a CNF formula $\Psi$, which is expected to be obtained by substituting the universally quantified variables in a QBF with their corresponding Herbrand functions (see Section~\ref{sec:preliminaries}). Performing this substitution directly in a zero-knowledge (ZK) setting is inherently difficult due to the complexity of evaluating Boolean functions under substitution. To address this challenge, we avoid direct substitution by leveraging the structure of CNF and instead verify that the target formula $\psi \land \psi_H$ and $\psi_H = \Tau(H)$. 
Here $\psi$ is the propositional matrix of the original QBF and is publicly known, while $\psi_H$ is a private witness representing the Tseitin CNF encoding of the Herbrand function $H$ via the transformation $\Tau$. 

Our goal now is to check the correctness of the CNF $\psi_H$ given a committed list of Herbrand function tuples. Recall that each tuple $(x_i, \ell^a_i, \ell^b_i) \in H$ specifies a Boolean constraint of the form $x_i \iff \ell^a_i \land \ell^b_i$. Applying the Tseitin transformation, this constraint can be equivalently encoded in CNF as the conjunction of three clauses: $TC^1_i = (\lnot x_i \lor \ell^a_i)$, $TC^2_i = (\lnot x_i \lor \ell^a_i)$, and $TC^3_i = (x_i \lor \lnot \ell^a_i \lor \lnot \ell^b_i)$. Prover constructs the witness CNF as
$
\psi_H = \bigwedge_i \left( TC^1_i \land TC^2_i \land TC^3_i \right),
$
and commits to it using a clause-level commitment scheme. To prove that each clause in $\psi_H$ correctly represents its corresponding gate in $H$, we can then leverage the $\Func[ZKLiteral]$ functionality to demonstrate membership of the expected literals in each clause that can be achieved by $\Func[Clause]$. 

\subsection{Refinement for Skolem functions} 
When proving that a QBF $\Psi = Q.\psi$ is true (i.e., satisfiable), the strategy involves concretizing each existentially quantified variable with a corresponding Boolean function, unlike in the Herbrandization case, which targets a subset of universally quantified variables. 
Nevertheless, we can adopt the same structural approach by encoding Skolem functions as a list \( S = \{(x_i, \ell^a_i, \ell^b_i)\} \), analogous to the representation used for Herbrand functions. To ensure completeness and correctness of the strategy, we also prove that every existential variable is assigned by checking if \( \varexists \subseteq \{x_i\} \) holds.

\section{Implementation and Evaluation}
\label{sec:eval}
We implement two protocols for verifying QBF evaluation in ZK: \zkqres, which is based on validating \qres proofs, and \zkws, which is based on validating winning strategies. In this section, we focus on presenting results for false QBFs, as they constitute the majority of instances for which we can obtain the certificates (\qres proofs and Herbrand functions). Results for true QBFs are discussed in Appendix~\Cref{sec:trueqbf}.

\subsection{Setup}
\paragraph{Implementation dependency.}
We implement our protocols using the EMP-toolkit~\cite{emp-toolkit} as the backend for interactive zero-knowledge proofs, and incorporate the open-source implementation of ZKUNSAT~\cite{zkunsat}. 
We solve QBFs and collect Q-Resolution proofs by DepQBF and QRPcheck\cite{niemetz2012resolution} and Skolem/Herbrand functions from CAQE~\cite{rabe2015caqe}\footnote{ We use earlier versions of DepQBF and CAQE, as recent releases do not support proof or certificate generation.}. 

\paragraph{Testbed.}
We perform our evaluation of our protocols on AWS {\tt i4i.16xlarge} instances, each equipped with 64 vCPUs, 512 GB RAM, and 50 Gbps inter-instance bandwidth between the prover and verifier, unless stated otherwise. High-memory instances are chosen due to the substantial memory demands of our largest benchmarks. 

We run the QBF solvers on compute nodes equipped with two Intel Xeon Gold 6148 CPUs (40 cores, 2.4 GHz) and 202 GB of RAM. This configuration is also used to generate and preprocess certificates, i.e., \qres proofs or winning strategies.

\paragraph{Benchmarks from QBFEVAL.}\label{sec:\QBFEVAL}
We use QBF formulae from the \QBFEVAL benchmark suite~\cite{QBFEVAL2023}, a widely adopted collection for evaluating solvers in QBF reasoning. We evaluate \zkqres and \zkws against the 2007 and 2023 editions\footnote{ The certified solver track was discontinued after 2016. To the best of our knowledge, \QBFEVAL'07 is the latest available benchmark suite that includes the track requiring certificates}. 
We present results for \QBFEVAL'07 in this section and include detailed results of \QBFEVAL'23 in the appendix\footnote{ Since 2023 benchmarks are not designed to evaluate certificate generation, we are only able to obtain proofs within our timeout configuration (described in Appendix~\Cref{sec:qbfeval23}) for 25 instances from \QBFEVAL'23.}. 
We evaluate \zkqres on 351 and \zkws on 322 instances from \QBFEVAL'07.
The remaining instances are excluded due to solver timeouts, proof preprocessing failures, or certificate sizes exceeding 25 MB\footnote {We set a 25MB limit, as the running time for larger proofs is estimated to exceed one hour. These instances are therefore excluded due to our resource constraints.}. We describe the details of the instance selection procedure in the Appendix~\Cref{sec:benchmarkselection}.

\subsubsection*{Benchmarks from real-world applications} We generate benchmark instances that encode the real-world applications introduced in Section~\ref{sec:intro}. The formulae are selected using the same procedure employed to obtain instances from the \QBFEVAL'07 benchmark set in our evaluation pipeline.

\noindent
\underline{PEC.}
The PEC benchmarks encode partial equivalence checking problems as QBFs. We use the benchmark suite provided in~\cite{gitina2013equivalence}. 

\noindent
\underline{C-PLAN.}
We generate QBF encodings for Blocks World planning problems using Q-Planner~\cite{qplanner2024}. Specifically, we extract five instances in which the plan length is strictly less than 5.

\noindent
\underline{BBC.}
We use benchmarks for BBC with QBFs provided by~\cite{seidl2012qbf2epr}. These formulae encode instances of black-box bounded model checking, a setting in which the internal structure of specific modules is unknown. Solving such QBF encodings is generally considered a challenging task. 

\subsection{Evaluation on \QBFEVAL}
\paragraph{\zkqres.} 
We evaluate the communication cost and prover running time of \zkqres for verifying \qres proofs across benchmark instances. Out of 351 instances, \zkqres successfully verifies 328 in ZK within 1.3 hours, while running out of memory for the rest of the instances. About 72\% of instances are verified within 100 seconds. The cumulative proportion of verified instances and the distribution of verification time are shown in~\Cref{fig:verified_dist}.

\begin{figure}[h]
    \centering
    \includegraphics[width=\linewidth]{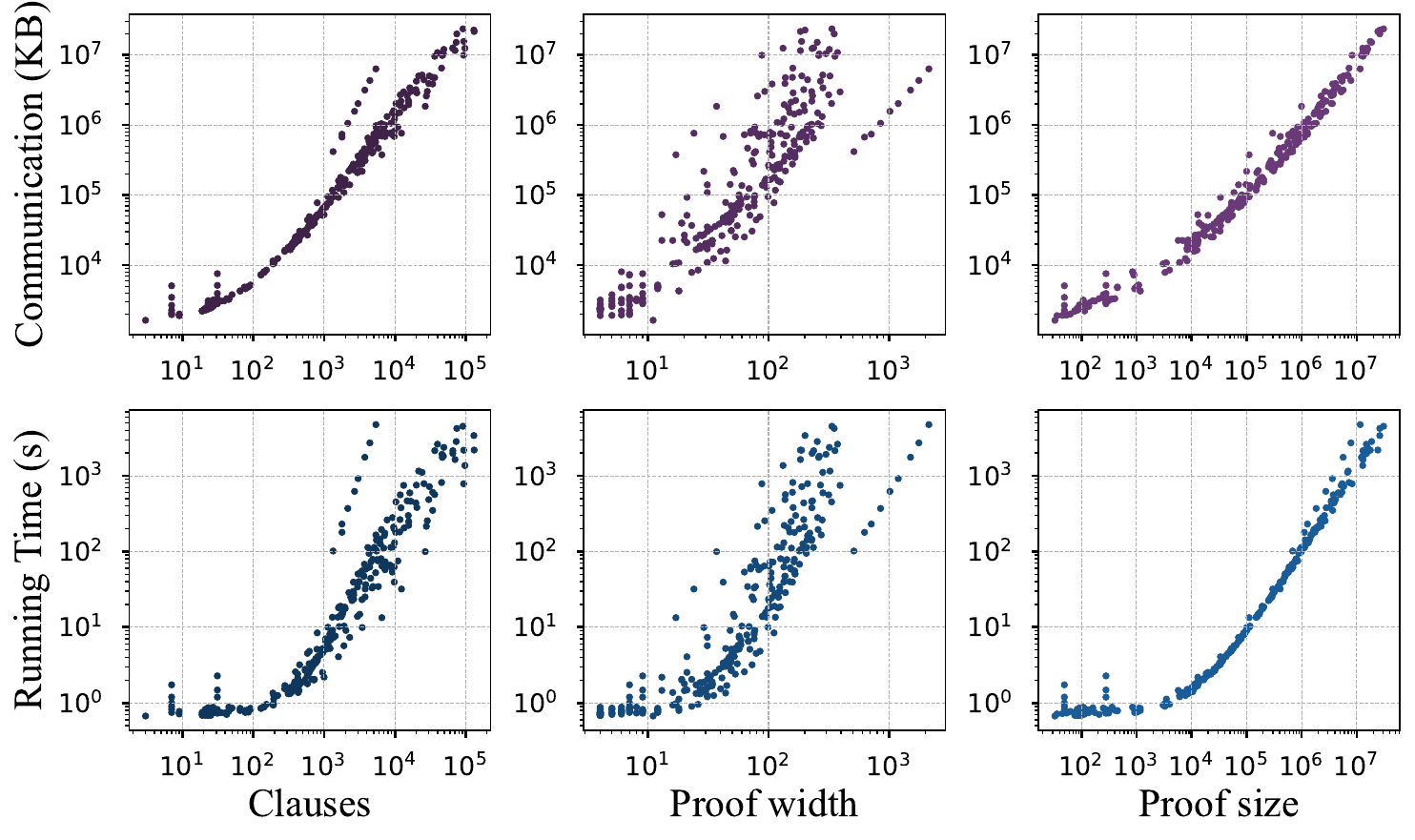}
    \caption{{\bf Communication and time cost of \zkqres.} Results are sorted by the number of clauses, proof width, and their product (which approximates the total proof size). The results indicate that the protocol's performance on \qres proofs scales closely with this product.}
    \label{fig:qrp_main}
\end{figure}

To understand the factors affecting the efficiency of our protocols, the results in Figure~\ref{fig:qrp_main} are sorted by the number of clauses, proof width, and their product, which approximates the overall proof size. 
The results indicate that the protocol's performance on \qres proofs scales relatively linearly with the proof size. 
Our evaluation shows that our protocol can verify instances with up to 2,000 \qres steps and proof width as large as 2,000. 
When the proof width is reduced to around 300, the system can handle proofs with up to 87,000 steps in a similar time frame. 

\paragraph{\zkws.} 
We evaluate the communication cost and prover running time of \zkws for verifying Herbrand function benchmarks.
Out of 322 instances, \zkws successfully verifies 283 of them within about half an hour, while it runs out of memory for the rest of the instances. 
The largest instance yields Herbrand functions of size up to 33k, with a proof width of approximately 315. About 82\% instances are verified within 100 seconds. The cumulative proportion of verified instances and the distribution of verification time are also shown in~\Cref{fig:verified_dist}.

\begin{figure*}[h]
    \centering
    \includegraphics[width=0.8\linewidth]{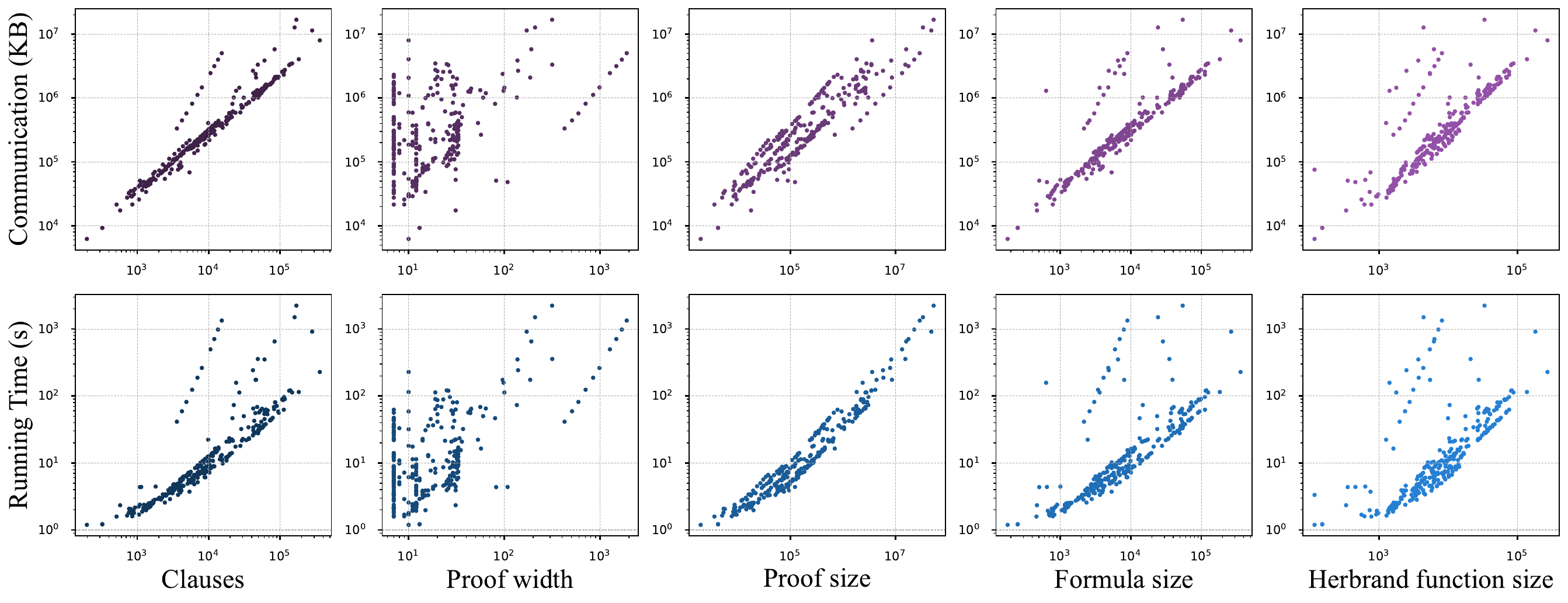}
    \caption{\footnotesize {\bf Communication cost and prover running time for Herbrand function benchmarks.} Results are sorted by number of clauses, proof width, and their product, reflecting the approximate proof size. Also shown are the size of CNFs resulting from Herbrandization and the size of the Herbrand functions themselves. Both communication and running time correlate most closely with the product of clause count and width, consistent with theoretical expectations of ZKUNSAT’s complexity.}\vspace{-4mm}
    \label{fig:herbrand_sorted}
\end{figure*}

Figure~\ref{fig:herbrand_sorted} presents the results, sorted by the number of clauses, proof width, proof size (i.e., the product of the number of clauses and proof width), the size of the CNF produced by Herbrandization, and the number of assignments in each Herbrand function. As the figure indicates, both communication and time costs scale approximately proportionally with the proof size, consistent with ZKUNSAT performance results.

\begin{figure}[h]
    \centering
    \includegraphics[width=\linewidth]{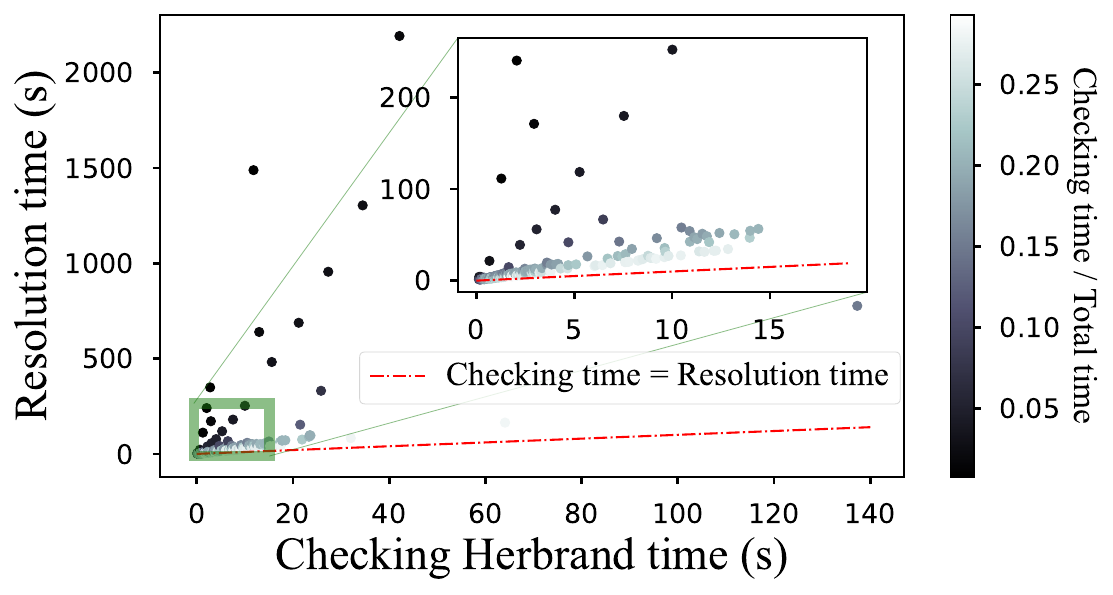}
    \caption{{\bf Decomposition of the verification time of \zkws.} The time costs are split into two components: (1) the time for checking Herbrandization correctness, which includes both well-formedness and correctness of the corresponding CNF encoding; and (2) the time for ZKUNSAT to verify UNSAT of the CNF produced by Herbrandization. Darker points indicate a larger proportion of time spent in ZKUNSAT, with the red line marking equal contribution from both components. The results show that ZKUNSAT is the primary performance bottleneck in \zkws.}
    \label{fig:cvsr} \vspace{-3mm}
\end{figure}

To understand the primary cost source and performance bottleneck, we break down the cost components involved in verifying Herbrand functions, as shown in Figure~\ref{fig:cvsr}. The results indicate that the unsatisfiability check remains the primary bottleneck in the overall verification process.

\paragraph{Comparison between \zkqres and \zkws.} We do not find an absolute advantage of one approach across all instances. For example, for some instances, \zkqres takes only 32 seconds, while it takes 2K seconds for \zkws. On the other hand, the other instance that takes 
4K seconds via \zkqres, while it takes only 22 seconds when using Herbrand functions with \zkws. We list the results in Figure~\ref{fig:verified_dist} by comparing the distribution of verification time using different approaches.

\subsection{Evaluation on Real-World Application}
We also evaluate our protocols on QBF instances encoding real-world applications. The detailed results are listed in Appendix~\Cref{tab:pec,tab:Q-Planner}, and~\Cref{fig:biu}.

\paragraph{PEC.} Our protocol successfully verifies a set of real-world circuit equivalence instances within 160 seconds using \qres. In contrast, the same instances can be verified in just 8 seconds using Herbrandization, highlighting the practical efficiency gains offered by our approach.

\paragraph{C-PLAN.} For planning problems generated from the Q-Planner examples, \zkqres successfully verifies the QBF instances encoding a blocks world like problem with step-length constraints less than 5 within 160 seconds. In contrast, \zkws can not verify instances with step-lengths greater than 3 due to the prohibitive size of the certificates.

\paragraph{BBC.} The largest instance of the BBC problem that can be verified by \zkws has proof width 316 and involves 48,802 clauses. \zkws takes 29 seconds to verify the correctness of Herbrandization. The dominant cost, however, remains in the ZKUNSAT phase, which verifies the unsatisfiability of Herbranded QBF in about 300 seconds.

\subsection{ZKUNSAT Optimization}
We utilize the \textsc{ZKUNSAT} framework to prove the unsatisfiability of Herbrandized or Skolemized QBFs. The performance of \textsc{ZKUNSAT} depends linearly on the clause width, defined as the maximum number of literals in any clause appearing in either the formula or the resolution proof. To ensure ZK, all clauses are padded to this maximum width, regardless of their actual size.

\begin{figure}[!]
    \centering
    \includegraphics[width=0.7\linewidth]{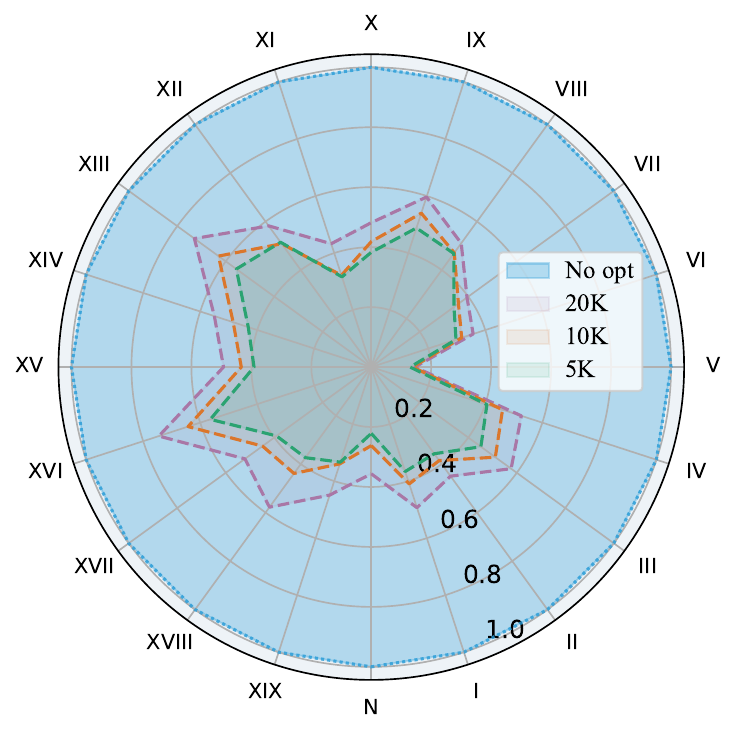}
    \caption{{\bf Running time for instance verification via Herbrandization w/wo clause partitioning optimization.} Each radius represents a QBF instance. We present the ratio of verification time after applying our optimization scheme with 5K, 10K, and 20K buckets, respectively. The results show that, for most instances, the time cost is reduced by half when using 10K buckets. For certain instances, the optimization achieves up to a 90\% improvement with 5K buckets.}
    \label{fig:optimization_running_time}
\end{figure}

We find that such padding leads to significant waste when proving knowledge of Herbrand/Skolem functions. In both the formula and the proof derived from Herbrandization/Skolemization, only a small fraction of the clauses have large literal width, while the majority of the clauses are narrow.  To mitigate this inefficiency, we introduce a clause partitioning strategy that reduces padding waste while preserving soundness. The key idea is to split the proof into different buckets, each corresponding to a subset of clauses with similar width. Such partitioning allows us to process narrower clauses without incurring the padding cost of the widest ones. The steps of the procedure are shown in Appendix~\Cref{alg:bucket}. While this reveals a part of the information, this structural leakage reduces padding overhead substantially. Shown in Figure~\ref{fig:optimization_running_time}, the performance improves by approximately 50\% when the bucket size is set to 10K.

\section{Information Leakage - Efficiency Trade-off:}

In this section, we present the information leakage - efficiency trade-off that our protocol offers. We begin by summarizing the information leaked by our protocols.

\paragraph{\zkqres (false QBFs).} We reveal upper bounds on the Q-Resolution proof width, on the number of Q-Resolution steps, and on the maximum number of literals removed in any \forallrmv.

\paragraph{\zkws (false QBFs).} We reveal upper bounds on the number of $\land$-gates in the AIG for the Herbrand function and, in ZKUNSAT (for proving Herbrandization is unsatisfiable), upper bounds on resolution proof width and length.

\paragraph{Mitigations and cost.} Leakage can be reduced by padding while preserving soundness: re-deriving clauses to mask true proof length; using higher-degree polynomials to obscure proof width and the maximum literals removed per step; and introducing unused variables in the Herbrand/Skolem function to hide its true size. These defenses incur overhead, reflecting an inherent efficiency trade-off.

\paragraph{Hardness.} The hardness of finding QBF proofs given such upper bounds is unknown. However, when bounds are set to natural worst-case values implied by the input QBF and its variable count, search remains at least as hard as without any auxiliary information.

\section{Acknowledgements:}
We thank Arijit Shaw and Anwar Hithnawi for many useful discussions, and Martina Seidl for providing the QBFEVAL 2007 benchmarks. We acknowledge the support of the Natural Sciences and Engineering Research Council of Canada (NSERC), funding reference number [RGPIN-2024-05956]. Part of the computational work for this article was performed on the Niagara supercomputer at the SciNet HPC Consortium. SciNet is
funded by Innovation, Science and Economic Development Canada; the Digital
Research Alliance of Canada; the Ontario Research Fund: Research Excellence;
and the University of Toronto. We also acknowledge CloudLab\cite{cloudlab} for providing the research infrastructure for our artifact evaluation process.

\bibliographystyle{IEEEtran}
\bibliography{abbrev3, crypto, bib}

\ifsubmission
\else
\appendix
\fi

\section{Proving True PCNF in ZK via \qres}
\label{sec:cube}
A QBF $\Psi = \Pi\psi$ in PCNF, is true if and only if the empty cube is derivable by Q-cube resolution and the model generation rule \cite{Giunchiglia2006clause/term}. We now introduce Q-cube resolution proofs and the model generation rule.

\noindent \textbf{Cubes.} A cube is defined to be the conjunction of a set of literals. For example, $(a \land b \land \lnot c)$.

\noindent\textbf{Q-Cube Resolution Step.}
A Q-cube resolution step is the derivation of a non-contradictory cube through the resolution of 2 cubes over a universally quantified variable.

\noindent\textbf{Existential Reduction.}
An Existential reduction of a cube is the removal of all existential variables that do not preceed any universally quantified variable.

\noindent\textbf{Q-Cube Resolution Proof.}
A Q-cube resolution proof is the derivation of the empty cube from a set of initial cubes through the application of Q-cube resolution steps and existential reduction. The initial cubes of a QBF in PCNF are a subset of the cubes derived by the model generation rule.

\noindent\textbf{Model Generation Rule.}
A set of cubes $\Phi$ is said to be derivable by the model generation rule if all of the following are true:
\begin{itemize}
    \item For each cube $D \in \Phi$ and each clause $C \in \psi$, $D \cap C \ne \phi$ (where $\phi$ is the empty set) and $D$ is non-contradictory.
    \item The disjunction of all cubes in $\Phi$ is propositionally logically equivalent to $\psi$.
\end{itemize}

\noindent \textbf{Proving True QBFs in ZK.}
By switching the quantifiers in the Q-Res protocol, we can convert it from a Q-Res checking protocol into a Q-Cube-Res checking protocol. However, since the public QBF is in PCNF, as an additional step, the prover needs to convince the verifier that the set of initial cubes they are operating with are valid (i.e. derivable via model generation rule and the existential reduction of this derived cube). We call the initial cube before existential reduction the pre-initial cube.

\noindent \textbf{Proving Initial Cube Validity in ZK.}
Suppose the prover wants to prove that the pre-initial cube $s = \{l_1, \dots, l_n\}$ is valid. If $s$ was a valid pre-initial cube, then by definition of the model generation rule for each clause $C$ in the public QBF $\psi$, there must exist some $l_{C_s} \in C\cap s$. For each clause $C \in \psi$, the prover gathers the literal $l_{C_s}$, then the prover and verifier then invoke $\Func[ZKSet]$ to check the subset relation $\bigcup_{C\in \psi} l_{C_s} \subseteq s$ and $l_{C_s} \subseteq C$. Now, all that remains is to prove that $s$ is non-contradictory. After this the prover and verifier can use the existentially reduced version of $s$ as the initial cube for the Q-cube resolution.

\noindent \textbf{Proving Non-Contradictory (Non-Tautological) Cubes (Clauses) in ZK} To prove that the cube (clause) $s$ is non-contradictory (non-tautological), the prover commits another cube (clause) $s'$ such that $l \in s \iff \lnot l \in s'$ and proceeds to prove that the GCD of the two polynomials $GCD(\gamma(s)(X), \gamma(s')(X))=1$ by committing $p(X)$ and $q(X)$ such that $1 = p(X) *(\gamma(s)(X)) + q(X) *(\gamma(s')(X))$. 

 \section{Functionalties}

\begin{figure}[h]
      \begin{nffunc}{$\Func[FlexZKArray]$}

   \shortsection{Array initialization:}
      On receiving $({\sf Init}, N,[m_0],\ldots,$ $[m_{N-1}])$ from \Prv and \Ver, where
      $m_i\in\FF$, store the~$\{m_i\}$ and set $f:={\sf honest}$ and ignore subsequent initialization calls.

  \shortsection{Array read:}
  On receiving $({\sf Read}, \ell, d,t)$ from \Prv, and $({\sf Read}, t)$ from \Ver, where $d\in\FF$ and
  $\ell,t \in \NN$,
  send $[d]$ to each party. If $d \neq m_\ell$ or $t$ from both parties do not match or $\ell \geq t$ then
  set $f:={\sf cheating}$.
 
  \shortsection{Array check:} Upon receiving $({\sf check})$ from $\Ver$ do: If \Prv sends $({\sf cheat})$ then send {\sf cheating} to $\Ver$. If \Prv sends $({\sf continue})$ then send     $f$ to~$\Ver$.
     \end{nffunc}
     \caption{Functionality for append-only arrays in ZK.}
     \label{idfunc:roram}
 \end{figure}

\begin{figure}[!t]
      \begin{nfprot}{\Prot[Clause]}
\shortsection{Res~\cite{zkunsat}:} 
  \begin{enumerate}[leftmargin = 4mm]
      \item $\Prv$ locally computes $W_{0}(X), W_{1}(X)$ and $\ell_p$, such that $W_{ 0}(X)\cdot \gamma(C_0)(X) = \gamma(C_r)(X) \cdot (X+\epsilon(\ell_p))$ and $W_{1}(X)\cdot \gamma(C_1)(X) = \gamma(C_r)(X) \cdot (X +\epsilon(\neg \ell_p))$. Note that the degree of $W_0(X)$ and $W_1(X)$ are bounded by $w$.
      \item $\Prv$ locally computes $\rho(X) = X - \epsilon(\ell_p)$, of which the degree is bounded by $1$.
      \item Two parties use $\Func[ZK]$ to authenticate all $w+1$ polynomial coefficients in $W_0(X)$ and $W_1(X)$, and two polynomial coefficients in $\rho(X)$. As a result, two parties get $[W_0(X)]$, $[W_1(X)]$ and $[\rho(X)]$.
      \item Using \Func[ZK], two parties check that the highest coefficient in $[\rho(X)]$ is non-zero, this make sense that $[\rho(X)]$ has degree exactly 1.
      \item \Prv locally computes polynomial $\bar \rho (X) = \rho(1_\FF-X)$ and commits its $2$ coefficients to obtain $[\bar \rho(X)]$. Then two parties check that the committed coefficients satisfy $\bar \rho (X) = \rho (1_\FF-X)$.
\item Both parties send $({\sf PoPEqCheck}, X, ([W_0(X)], [\gamma(C_0)(X)]),$ $([\gamma(C_r)(X)], [\rho(X)]))$  to $\Func[ZK]$. 
\item Both parties send $({\sf PoPEqCheck}, X, ([W_1(X)], [\gamma(C_1)(X)]),$ $([\gamma(C_r)(X)], [\bar{\rho}(X)]))$  to $\Func[ZK]$. 
\item As an additional step for \qres or Q-Cube Res, Non-tautological clause checking is performed on $C_r$ as outlined in Appendix-\Cref{sec:cube}
  \end{enumerate}

      \end{nfprot} 
     \caption{Checking resolution in ZKUNSAT.}
     \label{protocol:clauseres}
 \end{figure}

\section{ZKUNSAT Optimization}
\label{alg:bucket}
\begin{algorithm}
\caption{Clause Bucketing and Memory Array Construction}
\begin{algorithmic}[1]
\State Choose a parameter $k$ that determines the number of clauses per bucket.
\State Partition the set of clauses into buckets $\{G_1, G_2, \dots\}$ such that each bucket $G_j$ contains $k$ clauses.
\For{each bucket $G_j$}
    \State Compute $w_j \gets \max\{\text{width}(C) \mid C \in G_j\}$
    \State Construct a clause memory array storing all clauses in $G_j$, each padded to width $w_j$
\EndFor
\end{algorithmic}
\label{alg:bucket-clause}
\end{algorithm}

\newpage 
\section{Proof of \Cref{thm:herbrand_condition}}
\label{prf:herbrand}
\begin{proof}
We prove both directions.

\noindent
\textbf{(Only if).} Assume $H$ is a well-formed Herbrand function. Then by definition:
\begin{itemize}
    \item Each output variable is uniquely defined and drawn from $\varforall \cup \varaux$, implying Condition (1).
    \item The dependency graph induced by $H$ is acyclic. Since $H$ is topologically ordered, each variable $x_i$ only depends on earlier ones, satisfying Condition (2).
    \item Well-formedness requires that every universal variable $x_i \in \varforall$ only depends on literals of variables that appear before it in the quantifier prefix. The recursive function $\dep(x_i)$ captures this dependency level. Thus, $\dep(x_i) \leq \getorder(x_i)$, satisfying Condition (3).
\end{itemize}
\noindent
\textbf{(If).} Assume Conditions (1), (2), and (3) hold.

\begin{itemize}
    \item Condition (1) ensures that $H$ defines a function: each $x_i$ is assigned exactly once and from a valid domain.
    
    \item Condition (2) ensures acyclicity: since $x_i$ depends only on variables $x_j$ with $j < i$, the dependency graph has no cycles. This allows topological evaluation and confirms that $H$ is well-founded.
    
    \item Condition (3) ensures prefix-consistency. The function $\dep(x_i)$ computes the highest-order dependency of $x_i$. If $\dep(x_i) \leq \getorder(x_i)$, then all literals used in defining $x_i$ come from earlier in the prefix. Thus, the universal variable $x_i$ does not depend on any variable that appears later in the quantifier prefix.
\end{itemize}
Hence, $H$ is a syntactically well-defined and prefix-consistent Herbrand function. This completes the proof.
\end{proof}

\section{Detailed Evaluation Results}
\subsection{QBFEVAL 2007 Instance Selection}
\label{sec:benchmarkselection}
We evaluate the performance of \zkqres on 351 out of 1,136 instances from the benchmark set. 
DepQBF produces complete Q-resolution proof traces for 398 false QBFs and 81 true QBFs within a 5-minute timeout. This timeout is chosen as DepQBF already generates very large traces (exceeding 10 GB) within this time frame. Among these instances, the proof preprocessing step times out on 47 false QBFs, resulting in 351 false QBFs and 81 true QBFs against which \zkqres can be evaluated.  

We evaluate the performance of \zkws on 405 out of 1,136 instances from the benchmark set. CAQE successfully generates Skolem or Herbrand function certificates for 425 instances—332 unsatisfiable and 93 satisfiable. We apply ABC~\cite{abc} to minimize the extracted strategies. ABC's circuit minimization times out on 10 unsatisfiable and 3 satisfiable instances. For the remaining cases, SAT solving and proof processing complete successfully on 322 unsatisfiable and 20 satisfiable instances, resulting in \zkws evaluation on 342 instances \footnote{\small For each QBF clause with $n \geq 2$ literals, skolemization introduces at least $n-1$ new variables, likely making to frequent timeouts in picosat and our resolution proof preprocessor.}.

\subsection{Detailed Evaluation on QBFEVAL 2023}
\label{sec:qbfeval23}
\begin{table}[t]
\centering
\scriptsize
\begin{tabular}{|c|c|c|c||c|c|c|}
\hline
\textbf{Instance Name} & \multicolumn{3}{c||}{\textbf{ZKWS}} & \multicolumn{3}{c|}{\textbf{ZKQRES}} \\
\cline{2-7}
& Time (s)  & Degree & \#Steps & Time (s) & Degree & QRES Steps \\
\hline
arbiter-05-comp-error01-qbf-hardness-depth-8 & 27.3519 & 76 & 5966 & 3.73424 & 41 & 458 \\
\hline
arbiter-06-comp-error01-qbf-hardness-depth-11 & 1375.63 & 346 & 51800 & \multicolumn{3}{c|}{--} \\
\hline
gttt\_2\_1\_001020\_4x4\_torus\_w & 8172.08 & 704 & 118329 &\multicolumn{3}{c|}{--} \\
\hline
gttt\_2\_2\_000111\_4x4\_torus\_w & 6033.31 & 644 & 95601 &  \multicolumn{3}{c|}{--} \\
\hline
gttt\_2\_2\_000111\_4x4\_w & 6690.66 & 680 & 97363 & \multicolumn{3}{c|}{--}\\
\hline
hein\_14\_5x5-07.pg & 7576.82 & 650 & 127124 & 438.279 & 107 & 38766 \\
\hline
neclaftp4001 & 9.14755 & 9 & 3716 &\multicolumn{3}{c|}{--}\\
\hline
ntrivil\_query71\_1344n & 7.85795 & 50 & 666 & 17.8503 & 102 & 1276 \\
\hline
W4-Umbrella\_tbm\_05.tex.moduleQ3.8S.000001 & 1695.84 & 1349 & 22621 & 3718.06 & 1355 & 12816 \\
\hline
W4-Umbrella\_tbm\_25.tex.moduleQ3.7S.000003 & 2021.18 & 1568 & 21876 & 5807.78 & 1570 & 15799 \\
\hline
W4-Umbrella\_tbm\_26.tex.moduleQ3.7S.000003 & 420.314 & 875 & 8014 & 1639.92 & 878 & 11519 \\
\hline
GuidanceService & \multicolumn{3}{c||}{--} & 68.9327 & 181 & 2337 \\
\hline
hein\_07\_4x4-07.pg &\multicolumn{3}{c||}{--}& 566.292 & 90 & 61943 \\
\hline
\end{tabular}
\caption{Evaluation on QBFEVAL 2023 Benchmarks.}
\label{tab:qbfeval2023}
\end{table}

\subsection{Detailed Evaluation on True QBFs.}
\label{sec:trueqbf}
\subsection{ Verifying Skolemization} 
\begin{figure}
    \centering
    \includegraphics[width=\linewidth]{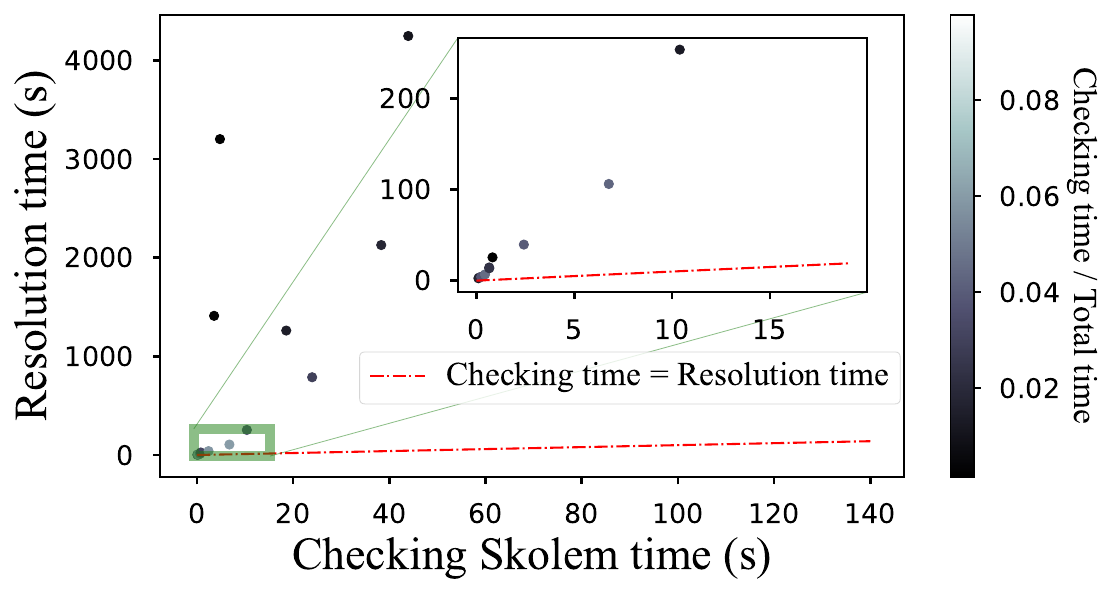}
    \caption{Comparison of the verification time for Skolemization against the time required for resolution-based unsatisfiability checking of the CNF formula produced by Skolemization. The results also indicate that resolution proof checking remains the primary bottleneck for Skolemization.}
    \label{fig:skolem_time_checking}
\end{figure}

We are unable to benchmark Skolemization functions on as many instances as we do for Herbrand functions for true QBFs, primarily due to the fact that the number of variables in the Skolemization is significantly higher than the number of variables in the QBF. This increases clause width in the proofs produced by picosat, making it hard for us to use ZKUNSAT. Figure~\ref{fig:skolem_time_checking} presents our evaluation results for the subset of instances we are able to verify.

\subsection{\qres for True QBFs via Cube Resolution} 
\begin{figure}
    \centering
    \includegraphics[width=\linewidth]{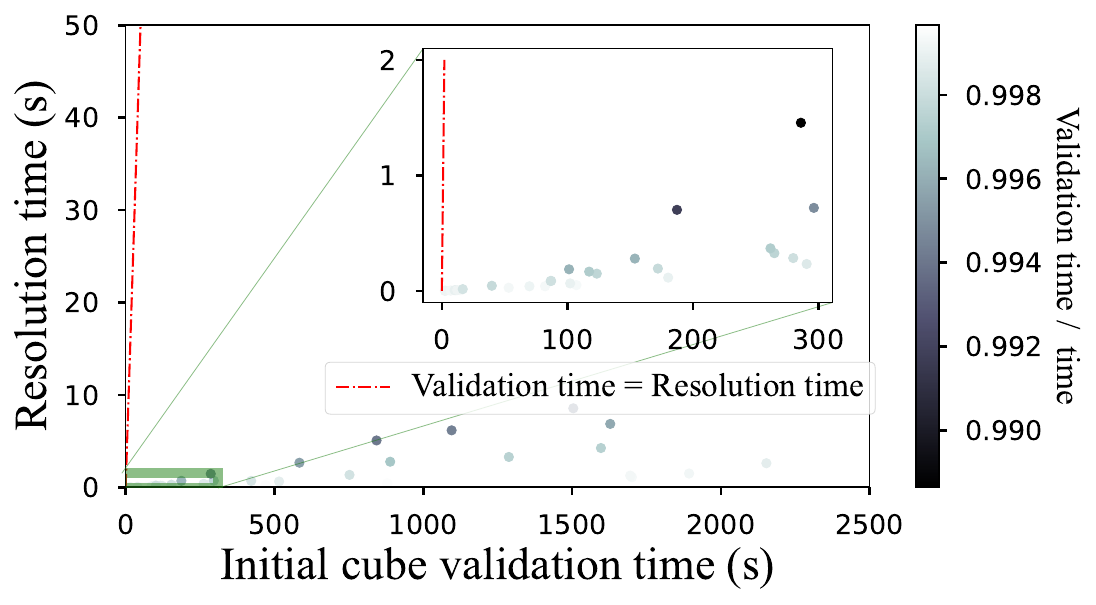}
    \caption{We compare the time cost of initial cube validation with that of cube resolution. The results show that initial cube validation is the primary bottleneck in verifying \qres proofs for true QBFs in ZK.}
    \label{fig:initial_cube_time_checking}
\end{figure}

\begin{figure}
    \centering
    \includegraphics[width=\linewidth]{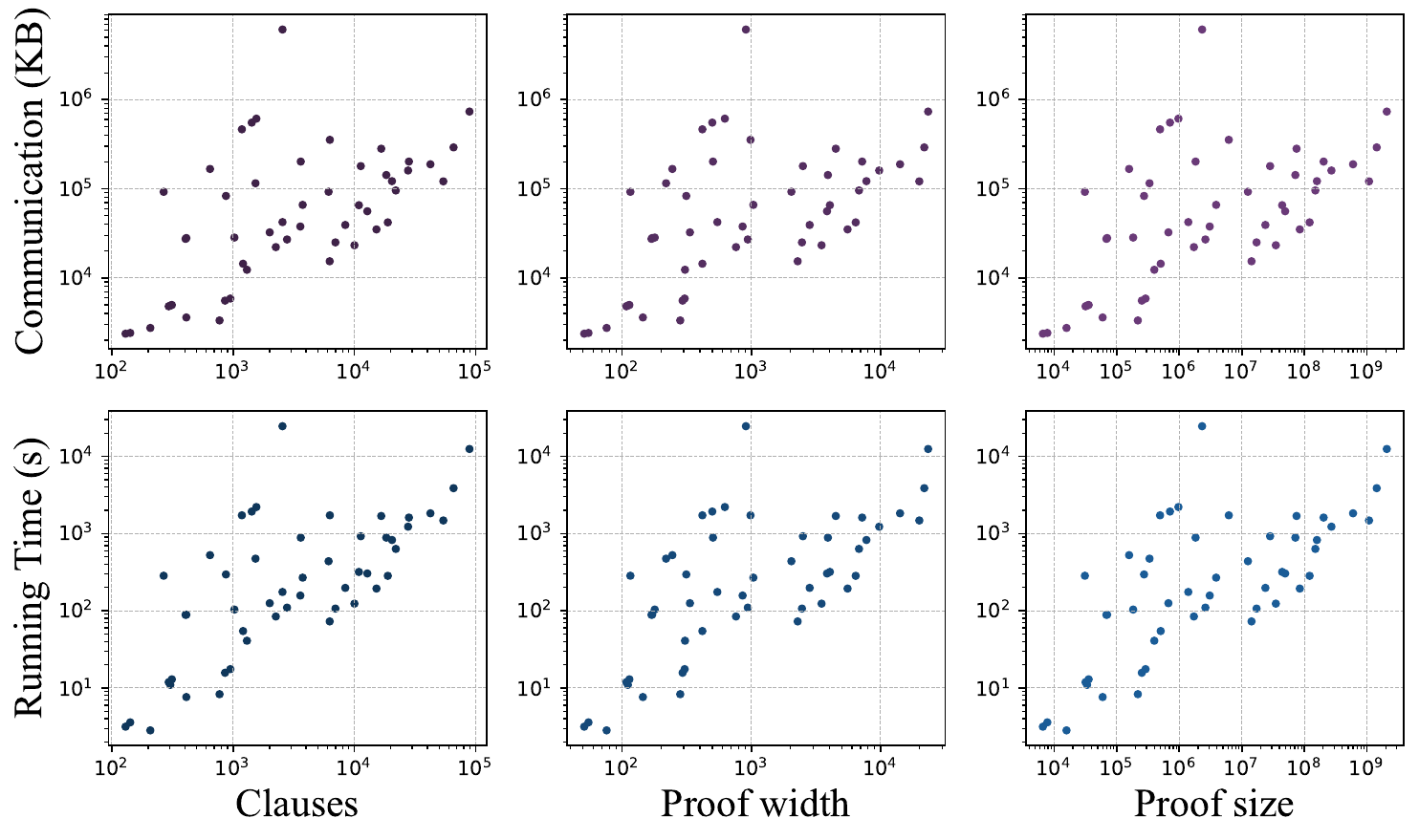}
    \caption{{\bf Communication and time cost of \zkqres for True QBFs.} Results are sorted by the number of clauses, proof width, and their product (which approximates the total proof size). Unlike \zkqres for false QBFs, the proportional relationship between the cost and the proof size is less apparent. This is because most of the cost arises from the validity checking of initial cubes.}
    \label{fig:qcube_time_checking}
\end{figure}

\subsection{Detailed Evaluation on Use Cases.}

\begin{table}[t]
\centering
\scriptsize
\begin{tabular}{|c|c|c||c|c|}
\hline
\textbf{Name} & \shortstack{\textbf{Herbrand}\\\textbf{Time (s)}} & \shortstack{\textbf{Herbrand}\\\textbf{Comm. (MB)}} & \shortstack{\textbf{ZKQRES}\\\textbf{Time (s)}} & \shortstack{\textbf{ZKQRES}\\\textbf{Comm. (MB)}} \\
\hline
z4ml.blif\_0.10\_0.20\_0\_1\_inp\_exact & 1.18 & 8.84 & 0.85 & 4.09 \\ \hline
z4ml.blif\_0.10\_0.20\_0\_1\_out\_exact & -- & -- & 0.71 & 2.13 \\ 
\hline
z4ml.blif\_0.10\_1.00\_0\_1\_inp\_exact & 1.17 & 8.84 & 0.79 & 4.09 \\ \hline
z4ml.blif\_0.10\_1.00\_0\_1\_out\_exact & 1.17 & 5.93 & 0.75 & 2.13 \\  \hline
comp.blif\_0.10\_0.20\_0\_1\_out\_exact & 1.56 & 24.57 & 1.54 & 15.03 \\  \hline
C499.blif\_0.10\_0.20\_0\_1\_out\_exact & 2.14 & 51.68 & 0.90 & 7.52 \\  \hline
C499.blif\_0.10\_1.00\_0\_1\_out\_exact & -- & -- & 0.96 & 8.12 \\  
\hline
C499.blif\_0.10\_1.00\_0\_1\_inp\_exact & -- & -- & 1.35 & 10.98 \\  
\hline
C880.blif\_0.10\_1.00\_0\_1\_out\_exact & 2.34 & 62.91 & -- & -- \\  
\hline
C880.blif\_0.10\_1.00\_0\_1\_inp\_exact & 3.23 & 72.93 & -- & -- \\  
\hline
term1.blif\_0.10\_0.20\_0\_1\_out\_exact & 2.74 & 72.09 & -- & -- \\  
\hline
term1.blif\_0.10\_1.00\_0\_1\_out\_exact & 3.00 & 77.27 & -- & -- \\  
\hline
term1.blif\_0.10\_1.00\_0\_1\_inp\_exact & 3.23 & 80.77 & -- & -- \\  
\hline
C6288.blif\_0.10\_0.20\_0\_1\_out\_exact & 7.70 & 326.57 & 55.49 & 387.23 \\  \hline
C6288.blif\_0.10\_1.00\_0\_1\_out\_exact & 8.65 & 371.79 & 165.72 & 933.17 \\  \hline
C5315.blif\_0.10\_1.00\_0\_1\_out\_exact & -- & -- & 0.73 & 2.96 \\  
\hline
C432.blif\_0.10\_1.00\_0\_1\_out\_exact & -- & -- & 1.60 & 16.25 \\  
\hline
comp.blif\_0.10\_1.00\_0\_1\_inp\_exact & -- & -- & 2.14 & 25.15 \\  
\hline
\end{tabular}
\caption{ZK Validation: PEC Benchmarks}
\label{tab:pec}
\end{table}

\begin{figure}[h]
    \centering
    \includegraphics[width=\linewidth]{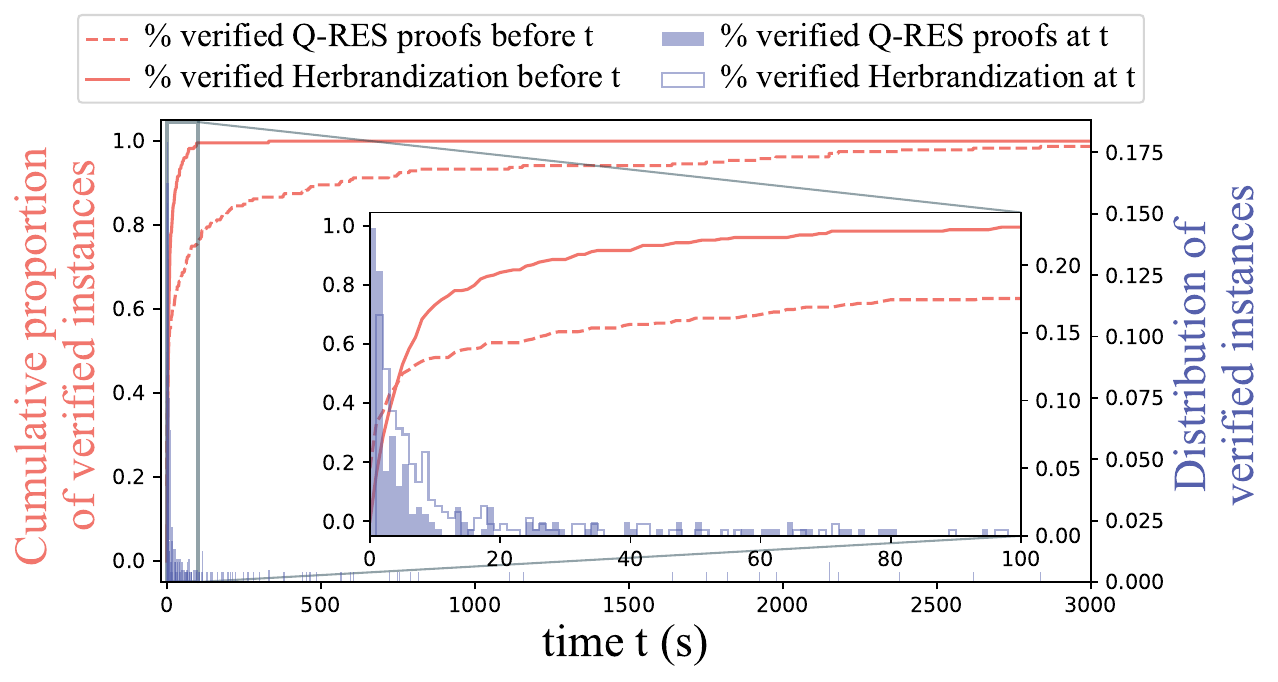}
    \caption{\footnotesize {\bf Evaluation results on BBC benchmarks.}We present the cumulative fraction of BBC QBF instances successfully verified via Q-RES proofs and winning strategies within a given time threshold (left $Y$-axis), as well as the fraction of instances verified around each time point (right $Y$-axis)}
    \label{fig:biu}
\end{figure}

\vspace{1em}

\begin{table}[t]
\centering
\scriptsize
\begin{tabular}{|c|c|c||c|c|}
        \hline
        \textbf{Name} & \shortstack{\textbf{Herbrand}\\\textbf{Time (s)}} & \shortstack{\textbf{Herbrand}\\\textbf{Comm. (MB)}} & \shortstack{\textbf{ZKQRES}\\\textbf{Time (s)}} & \shortstack{\textbf{ZKQRES}\\\textbf{Comm. (MB)}} \\
        \hline
        prob01, plan length = 1 & 6.01 & 13.82 & 2.20 & 16.56 \\
        \hline 
        prob01, plan length = 2 & 18.55 & 72.60 & 4.15 & 49.86 \\ \hline
        prob01, plan length = 3 & 153.28 & 599.54 & 155.99 & 1238.06 \\
        \hline
        prob01, plan length = 4 & 915.44 & 2879.52 & -- & -- \\ \hline
        prob01, plan length = 5 & 1105.94 & 8373.44 & -- & -- \\
     \hline
    \end{tabular}
    \caption{ZK QRP validation: Q-Planner Benchmarks (Blocks Q-Planner)}
    \label{tab:Q-Planner}
\end{table}

\end{document}